\documentclass[12pt,preprint]{aastex}

\usepackage{amsmath}

\newcommand\grm{{\tt grmonty}}
\newcommand\ibothros{{\tt ibothros}}

\newcommand\sC{{\mathcal C}}

\newcommand\<{{\langle}}
\renewcommand\>{{\rangle}}
\newcommand\eno{{\langle\epsilon\rangle}}

\def\gdet{{\sqrt{-g}}}

\newcommand\eps{{\epsilon}}

\begin{document}

\title{{\tt grmonty}: a Monte Carlo Code for Relativistic Radiative Transport}

\author{Joshua C. Dolence, Charles F. Gammie\altaffilmark{1}}
\affil{Astronomy Department, University of Illinois, Urbana, IL 61801}
\altaffiltext{1}{Physics Department, University of Illinois, Urbana, IL 61801}
\author{Monika Mo{\'s}cibrodzka}
\affil{Physics Department, University of Illinois, Urbana, IL 61801}
\author{Po Kin Leung}
\affil{Astronomy Department, University of Illinois, Urbana, IL 61801}
\email{dolence2@astro.illinois.edu}

\shortauthors{Dolence et al.}  \shorttitle{grmonty}

\begin{abstract}

We describe a Monte Carlo radiative transport code intended for calculating
spectra of hot, optically thin plasmas in full general relativity.  The
version we describe here is designed to model hot accretion flows in the
Kerr metric and therefore incorporates synchrotron emission and
absorption, and Compton scattering.  The code can be readily
generalized, however, to account for other radiative processes and an
arbitrary spacetime.  We describe a suite of test problems, and
demonstrate the expected $N^{-1/2}$ convergence rate, where $N$ is the
number of Monte Carlo samples.  Finally we illustrate the capabilities
of the code with a model calculation, a spectrum of the slowly accreting
black hole Sgr A* based on data provided by a numerical general
relativistic MHD model of the accreting plasma.

\end{abstract}


\section{Introduction}

There is wide interest in calculating the emergent radiation from
relativistic astrophysical sources, including accreting black holes,
accreting neutron stars, and relativistic blast waves.  A variety of
methods for solving the radiative transfer problem in these sources have
been developed over the last few decades \citep[e.g.][]{PSS, GW84, HW91,
CK92, CBR93, S95, PS96, Z96, DWB97, BL01, SKH06, noble07, Wu08}, some
based on Monte Carlo schemes.  Few of these schemes take full account of
relativistic effects in the source, however, and this is crucial in
estimating the spectra of hot plasma deep in a gravitational potential
well, or highly relativistic blast waves.  Monte Carlo transport of
radiation in accretion flows around compact objects has been considered
by \citep[e.g.][]{SK09,Sc06, YZ05, BJL03, BL01, LT99, AB96, S95}.  Among
others, \citet{Cu01, MB99, H97, GW84, PSS} give more general discussions
of Monte Carlo radiative transfer techniques.

We were motivated by efforts to model the radio source and black hole
candidate Sgr~A*.  Our interest in this source drove us to develop a
numerical scheme that could accurately calculate spectra of a
relativistic source in which the plasma properties (velocity, density,
magnetic field strength, and temperature) were specified by a separate
model---that is, sources in which radiation plays a negligible role in
the dynamics and energetics.  The result, a Monte Carlo scheme called
\grm, is described in this paper.  The spirit of our calculation is to
obtain an accurate spectrum with as few approximations as possible.  To
this end we treat Compton scattering with no expansions in $v/c$, and
allow for general angle-dependent emission and absorption (we specialize
to thermal synchrotron in this work).

In designing \grm\ our philosophy has been to maximize the physical
transparency and minimize the length of the code, occasionally at the
cost of reduced performance.  Sometimes simplicity and efficiency are in
harmony.  We chose to directly integrate the geodesic equation rather
than using a scheme that relies on integrability of geodesics in the
Kerr metric.  We will show that for radiative transfer problems where
many points are required along each geodesic, direct integration is not
only simpler and easier to modify, but also faster.

Our paper is organized as follows.  In \S\ref{MANUFACTURE} we describe
how we sample emission, and in \S\ref{GEODESIC} we describe how we track
photons along geodesics.  Evolution of superphoton weights under
absorption is described in \S\ref{ABSORB}, and sampling of scattered
photons is discussed in \S\ref{SCATTER}.  Photons at large distance from
the source must be sampled and assembled into spectra; this is described
in \S\ref{CENSUS}.  The code has been extensively verified; we describe
tests in \S\ref{TESTS}.  \S\ref{SAMPLE} describes a sample calculation,
and \S\ref{SUMM} summarizes our results.

Throughout this paper we assume that there is an underlying model that
can be queried to supply the rest-mass density $\rho_0$, the internal
energy $u$, the four-velocity $u^\mu$, and magnetic field four-vector
$b^\mu$ for the radiating plasma.  Usually we expect the model to be
supplied by a numerical simulation in a coordinate basis $x^\mu$.  

\section{Manufacturing Superphotons}\label{MANUFACTURE}

Emission in \grm\ is treated by sampling the emitted photon field.  The
samples, here called ``superphotons'' (also ``photon packets''), have
weight $w$, coordinates $x^\mu$, and wave vector $k^\mu$.  The weight $w
\gg 1$ is a pure number that represents the ratio of photons to
superphotons: $dN = w dN_s$ ($N_s \equiv$ number of superphotons, $N
\equiv$ number of photons).  In our models the weight is a function of
the emitting plasma frame frequency $\nu$ and nothing else.  The
coordinates are typically in model units (e.g. for a black hole
accretion flow calculation, length unit $L = G M/c^2$), and the
components of $k^\mu$ are given in units of $m_e c^2$.

How should superphotons be distributed over $x^\mu$ and $k^\mu$?  The
initial superphoton momentum can be described in an orthonormal tetrad
basis $e^\mu_{(a)}$ that is attached to the plasma, so that $e^\mu_{(t)}
= u^\mu$ ($\mu$ is the coordinate index, and $(a)$ is the index
associated with the tetrad basis, raised and lowered using the Minkowski
metric).  In the tetrad basis $k^{(a)}$ is specified by frequency $\nu$
and spatial direction unit vector $\hat{\bf n}$ that is contained within
the solid angle $d\Omega$.  The probability distribution for
superphotons is then
\begin{equation}\label{PROBDIST}
\frac{1}{\gdet}\,
\frac{d N_s}{d^3x dt \, d\nu d\Omega} =
\frac{1}{w\gdet} \, \frac{d N}{d^3x dt \, d\nu d\Omega} =
\frac{1}{w} \frac{j_\nu}{h\nu}
\end{equation}
where $j_\nu$ is the emissivity (always defined in the plasma frame),
since $\gdet \, d^3x dt$ is invariant (meaning coordinate invariant).  
In a time interval $\Delta t$ we expect to create 
\begin{equation}
N_{s,tot} = \Delta t \int \; \gdet \; d^3x \; d\nu \; d\Omega
\frac{1}{w}\frac{j_\nu}{h\nu}
\end{equation}
superphotons over the entire model volume.  The total computational
effort is proportional to $N_{s,tot}$, so we control the computational
effort by scaling the weights.  

How should we distribute superphotons over the volume?  \grm\ subdivides
the model volume into volume elements (``zones'') of size $\Delta^3 x$
(e.g. in Boyer-Lindquist $t, r, \theta, \phi$ coordinates for the Kerr
metric, $\Delta r \Delta \theta \Delta \phi$).  For zone $i$ we expect
\begin{equation}
N_{s,i} = \Delta t \Delta^3x \gdet \int \; d\nu \; d\Omega \;
\frac{1}{w}\, \frac{j_\nu}{h\nu}
\end{equation}
to be created in time $\Delta t$.  We create $N_{s,i}$ superphotons at
the center of zone $i$.  Fractional $N_{s,i}$ are dealt with by
rejection sampling.

The momentum-space (wave vector) piece of the probability distribution
(\ref{PROBDIST}) can be sampled by techniques outlined below to give
$\nu$ and $\hat{\bf n}$.  With $x^\mu$, $\nu$ and $\hat{\bf n}$ in hand,
we can construct $k^{(a)}$ and, finally, transform to the coordinate basis
$e^\mu_{(a)} k^{(a)} = k^\mu$.

The remaining ingredients in the sampling procedure are the emissivity,
the orthonormal tetrads, and the sampling procedures for $\nu$ and
$\hat{\bf n}$.

\subsection{Emissivity}

\grm\ depends on the emissivity only through functions that specify
$j_\nu$ and $\int d\nu d\Omega j_\nu/\nu$, so it is straightforward to
include any emission/absorption process.

In our target problem the only source of superphotons is thermal
synchrotron emission at dimensionless temperature $\Theta_e \equiv k
T_e/(m_e c^2)$.  \citet{L09} show that, for $\Theta_e \gtrsim 0.5$, 
\begin{subequations}
\begin{align}
j_\nu(\nu,\theta) &\simeq \frac{\sqrt{2}\pi e^2 n_e\nu_s}{3 c
K_2(\Theta_e^{-1})}
\left(X^{1/2} + 2^{11/12} X^{1/6} \right)^2
\exp\left(-X^{1/3}\right) \\
X &\equiv \frac{\nu}{\nu_s} \label{eqn:nu_nus}\\
\nu_s &\equiv \frac{2}{9}
\left(\frac{e B}{2\pi m_e c}\right) \Theta_e^2 \sin\theta 
\end{align} 
\end{subequations}
where $K_2$ is the modified Bessel function of the second kind, $n_e$ is the number density of electrons, $B$ is the magnetic field strength, and $\theta$ is the angle between the wave vector and magnetic field.  For
large $\Theta_e$, $K_2(\Theta_e^{-1}) \simeq 2 \Theta_e^2$, but for
$\Theta_e \lesssim 1$ better agreement with the emissivities of
\citet{L09} is obtained if $K_2$ is evaluated directly.  Since the
emissivity must be evaluated many times, it is most efficient to
precompute $K_2(\Theta_e^{-1})$ at the beginning of the calculation and
store the results in a table.

\subsection{Orthonormal tetrads}

The wave vector sampling is done in an orthonormal tetrad attached to the
fluid.  We construct the orthonormal tetrad $e^\mu_{(a)}$ using
numerical Gram-Schmidt orthogonalization.  Here $\mu$ is the coordinate
index, and $(a)$ is the index associated with the tetrad basis, which is
raised and lowered using the Minkowski metric.

We set $e^\mu_{(0)} = u^\mu$ ($u^\mu \equiv$ plasma four-velocity), and
then use $b^\mu$, the magnetic field four-vector, as the first trial
vector (this is numerically convenient since we will want to orient wave vectors with respect to the magnetic field; if $b^\mu = 0$ then we use a default, radius-aligned, trial
vector).  Thus $e^\mu_{(1)} = {\rm NORM}\left(b^\mu - e^\mu_{(0)}
(e^\nu_{(0)} b_\nu)\right)$ (${\rm NORM}$: normalize).  The process is
repeated with additional trial vectors to create a full tetrad basis.

The tetrad-to-coordinate basis transformation is 
\begin{equation}\label{eqn:tet_to_coord} 
k^\mu = e^\mu_{(a)} k^{(a)} 
\end{equation}
and the coordinate to tetrad transformation is
\begin{equation}\label{eqn:coord_to_tet} 
k^{(a)} = e^{(a)}_\mu k^\mu.  
\end{equation}
With these transformations in hand, we can construct the superphoton
wave vector in the orthonormal frame and then transform it to the
coordinate basis.

\subsection{Wave vector sampling procedure}

\subsubsection{Photon energy}

Within zone $i$ superphotons are distributed over frequency according to 
the distribution
\begin{equation}\label{eqn:dns_zone}
\frac{dN_{s,i}}{d\ln\nu} = \Delta t \Delta^3 x \gdet \; \frac{1}{h\, w} \;
	\int d\Omega \; j_\nu.
\end{equation}
We sample the distribution only between a minimum and maximum frequency
$\nu_{min}$ and $\nu_{max}$.  These must be chosen so that no
significant emission is omitted from the final spectrum.  

We distribute superphotons over frequency by rejection sampling.  For
simplicity, we use a constant envelope function equal to the maximum of
Equation~\eqref{eqn:dns_zone} (for zone $i$). Thus we draw tentative
values uniformly in $\ln\nu$ from $\nu_{min}$ to $\nu_{max}$.  The
efficiency of the sampling procedure is the ratio of the areas under the
distribution and envelope, so if the distribution given in
Equation~\eqref{eqn:dns_zone} is sharply peaked this technique can be
inefficient.

In practice, we choose a tentative frequency $\nu_{0}=\exp(r_1
\ln{\nu_{max}/\nu_{min}} + \ln{\nu_{min}})$, where $r_1$ is drawn from a
uniform distribution on [0,1) (we use the Mersenne twister random number
generator from the GNU Scientific Library, hereafter GSL).  A second
number $r_2$ is drawn from [0,1) and the process is repeated until
\begin{equation} 
r_2 < \left.{\left. \frac{dN_{s,i}}{d\ln\nu}\right \vert}_{\nu_0}\right /
\mathrm{MAX}\left(\frac{dN_{s,i}}{d\ln\nu}\right).
\end{equation} 
The efficiency of this process is $\sim15\%$, but the cost is small
compared to the total cost of \grm.

\subsubsection{Photon direction}

The superphoton direction $\hat{\bf n}$ is described by polar
coordinates $\theta$ and $\phi$ in the tetrad frame, where $\theta$ is
the angle between the spatial part of the wave vector and the magnetic
field.  The colatitude $\theta$ is obtained by rejection sampling:  a
tentative value for $\theta$ is obtained by drawing $\mu =
\cos\theta$ from a uniform distribution on [-1,1), a second number
$r$ is drawn from a uniform distribution on [0,1), and $\theta$ is
accepted if
\begin{equation} 
r < \frac{j_\nu(\theta)}{j_\nu(\pi/2)}
\end{equation}
(this procedure is specific to the synchrotron emissivity).  The
efficiency of this scheme is problem dependent; for our target
application the efficiency is $\sim 65\%$.  Finally, $\phi$ is drawn
from a uniform distribution on [0,2$\pi$).

\subsubsection{Transformation to coordinate frame}

Once $\theta$, $\phi$, and $\eps=h\nu/m_ec^2$ are selected, the wave vector is 
completely specified in the orthonormal tetrad frame:
\begin{subequations}
\begin{align}
k^{(0)} &= \eps \\
k^{(1)} &= \eps\cos\theta \\
k^{(2)} &= \eps\sin\theta\cos\phi \\
k^{(3)} &= \eps\sin\theta\sin\phi,
\end{align}
\end{subequations}
and the wave vector in the coordinate frame is $k^\mu = e^\mu_{(a)} k^{(a)}$.

\section{Geodesic Integration}\label{GEODESIC}

General relativistic radiative transfer differs from conventional
radiative transfer in Minkowski space in that photon trajectories are no
longer trivial; photons move along geodesics.  Tracking geodesics is a
significant computational expense in {\tt grmonty}.

The governing equations for a photon trajectory are
\begin{equation}\label{eqn:geod_dxdlam}
\frac{d x^\alpha}{d\lambda} = k^\alpha
\end{equation} 
which defines $\lambda$, the affine parameter, the
geodesic equation
\begin{equation}\label{eqn:geod_dkdlam} 
\frac{d k^\alpha}{d\lambda} = -\Gamma^\alpha_{\mu\nu} k^\mu k^\nu
\end{equation} 
and the definition of the connection coefficients
\begin{equation}\label{eqn:connect} 
\Gamma^\alpha_{\mu\nu} = \frac{1}{2} g^{\alpha\gamma} \left(g_{\gamma\mu,\nu} + g_{\gamma\nu,\mu} - g_{\mu\nu,\gamma}\right)
\end{equation}
in a coordinate basis.  

We assume nothing about the metric, so it is easy to change coordinate
systems and even to extend the code to dynamical spacetimes.
Nevertheless, our main application---to black hole accretion flows---is
in the Kerr metric, where geodesics are integrable.  The four constants
of the motion are the energy-at-infinity $E$ (in Boyer-Lindquist
coordinates $t,r,\theta,\phi$, $E = -k_t$), the angular momentum $l =
k_\phi$, Carter's constant $Q = k_\theta^2 + k_\phi^2 \cot^2\theta - a^2
k_t^2 \cos^2\theta$ \citep[see][]{C68}, and the condition that
$k^\alpha$ be null: $k^\mu k_\mu = 0$, equivalent to the dispersion
relation for photons {\it in vacuo}: $\omega^2 = c^2 k^2$.  These four
constants of the motion can be used to quasi-analytically obtain $x^\mu$
and $k^\mu$ in terms of an initial (or final) position and wave vector
\citep[see, e.g.,][]{RB94,BD05,DA09}.  

The integrability of geodesics in the Kerr metric would appear to
provide an opportunity for significant computational economies.  We show
below, however, that direct integration of equations~\eqref{eqn:geod_dxdlam} 
and \eqref{eqn:geod_dkdlam} is not only simpler
and more flexible but also faster than at least one implementation of an
integral-based technique.

Which ODE integration algorithm is best for the geodesic equation?  If
only a few coordinate evaluations are required over the entire geodesic
then a high order scheme is optimal.  For example, we have found that
the embedded Runge-Kutta Prince-Dorman method available in GSL is fast
and accurate; it can easily be made to conserve the integrals of motion
to machine precision.  Many coordinate evaluations are required,
however, when integrating the equation of radiative transfer, as \grm\
does, along superphoton trajectories.  A second order scheme can then
provide the required accuracy at minimal cost.

Evaluating the connection coefficients is expensive, so we want to
choose a scheme that minimizes the number of evaluations.  The velocity
Verlet algorithm, which for the geodesic equation is
\begin{subequations}
\begin{align}
x^{\alpha}_{n+1}&=x^{\alpha}_n+k^{\alpha}_n\Delta\lambda+{{1}\over{2}}\left({{dk^{\alpha}}\over{d\lambda}}\right)_n(\Delta\lambda)^2 \\
k^{\alpha}_{n+1,p}&=k^{\alpha}_n+\left({{dk^{\alpha}}\over{d\lambda}}\right)_n\Delta\lambda \\
\left({{dk^{\alpha}}\over{d\lambda}}\right)_{n+1}&=-\Gamma^{\alpha}_{\mu\nu}(\mathbf{x}_{n+1})k^{\mu}_{n+1,p}k^{\nu}_{n+1,p}\label{eqn:geod_int_iter_begin} \\
k^{\alpha}_{n+1}&=k^{\alpha}_n +
{{1}\over{2}}\left(\left({{dk^{\alpha}}\over{d\lambda}}\right)_n +
\left({{dk^{\alpha}}\over{d\lambda}}\right)_{n+1}\right)(\Delta\lambda)
\label{eqn:geod_int_iter_end},
\end{align}
\end{subequations}
requires only one evaluation of the connection coefficients per step.
Accuracy can be improved by using the result of
equation~\eqref{eqn:geod_int_iter_end} to recompute the derivative
equation \eqref{eqn:geod_int_iter_begin} with
$k^\mu_{n+1,p}=k^\mu_{n+1}$ and then reevaluating
equation~\eqref{eqn:geod_int_iter_end}.  This process is repeated until
the change in the wave vector between estimates is below some tolerance.
In \grm\, we continue this iteration until the fractional change is less
than $10^{-3}$ (typically only once or twice).  This does not require
any additional evaluations of $\Gamma^{\alpha}_{\mu\nu}$.  Very rarely
we find that this iteration fails to converge, and then \grm\ defaults
to taking the step with a classical $4^{\rm th}$-order Runge-Kutta
technique.

\begin{figure}
\plotone{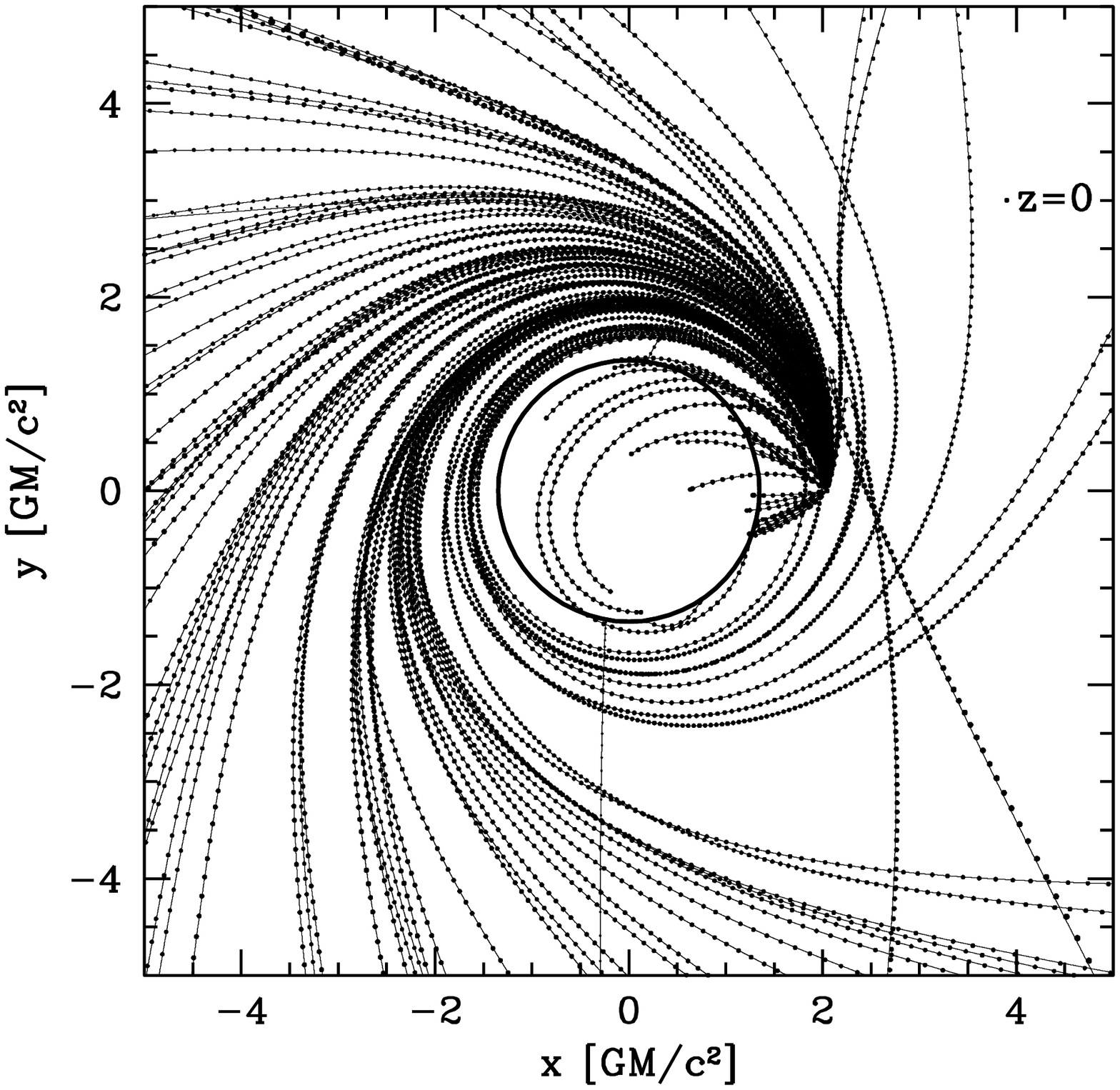}
\caption{Photon geodesics for isotropic emission from the rest frame of a fluid element 
in a marginally stable circular orbit around a Kerr black hole with $a/M=0.9375$. 
Results shown from \grm\ (points) and {\tt geokerr} (lines).  The point
size varies linearly with the z-coordinate.}
\label{fig:rf_geod}
\end{figure}

How fast and accurate is our geodesic integration scheme?  We propose
the following benchmark.  Consider a point on a direct, circular,
marginally stable orbit in the equatorial plane of a black hole with
spin $a/M = 1 - 2^{-4} = 0.9375$ that emits radiation isotropically in
its rest frame.  Sample the emitted photons (in a Monte Carlo sense; the
analytic circular orbit orthonormal tetrads available in \citet{B72} are
useful for constructing the initial wave vectors) and track them until
they cross the horizon or reach $r c^2/(G M) = 100$ ($r$ is the
Boyer-Lindquist or Kerr-Schild radial coordinate).
Figure~\ref{fig:rf_geod} shows as dots a representative sample of photon
geodesics from \grm\ in the coordinate frame, illustrating the effects
of relativistic beaming, lensing, and frame dragging. 

\begin{figure}
\plotone{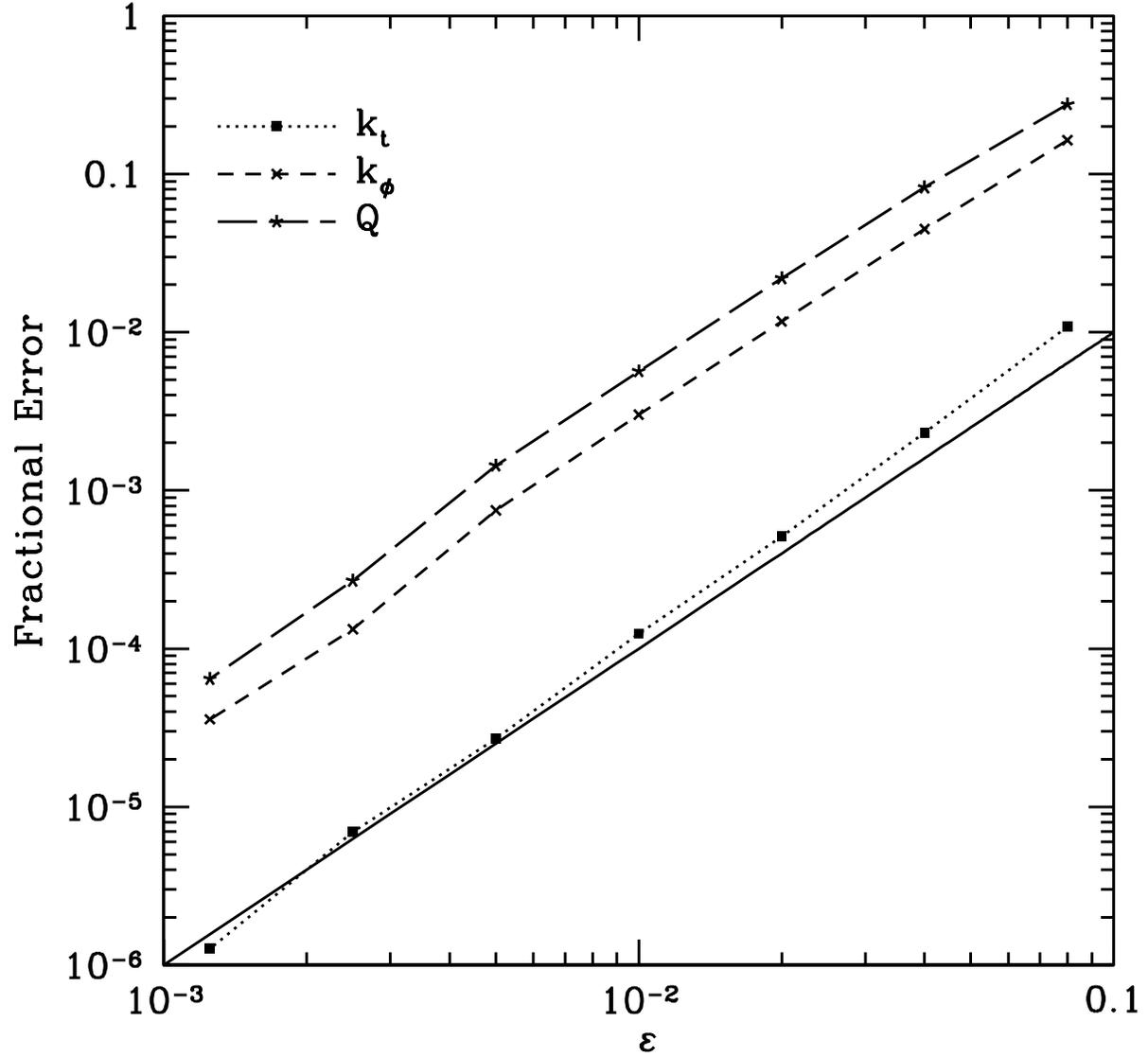}
\caption{Average fractional error in the conserved quantities $k_t$ and $k_{\phi}$ 
as a function of step size parameter $\varepsilon$.  The solid line is
$\varepsilon^2$, showing that \grm's geodesic integrator converges at 
second order.}
\label{fig:geod_cons}
\end{figure}

Second order convergence of the velocity Verlet integration scheme is
demonstrated in Figure~\ref{fig:geod_cons}, which plots the average
fractional error in $E$, $l$, and $Q$ as a function of a step-size parameter
$\varepsilon$.  We typically set $\varepsilon = 0.04$ as a compromise between
performance and accuracy; the average fractional errors at the end of the integrations are $\sim 2\times 10^{-3}$, $\sim 4\times 10^{-2}$, and $\sim 8\times 10^{-2}$ for $E$, $l$, and $Q$, respectively.  We have verified that this choice makes
geodesic tracking errors subdominant in the error budget for the overall
spectrum.

With $\varepsilon = 0.04$, \grm\ integrates $\sim 16,700\textrm{
geodesics sec}^{-1}$ on a single core of an Intel Xeon model E5430.  If
we use $4^{\rm th}$-order Runge-Kutta exclusively so that the error in
$E$, $l$, and $Q$ is $\sim$1000 times smaller, then the speed is $\sim
6,200\textrm{ geodesics sec}^{-1}$.  If we use the Runge-Kutta
Prince-Dorman method in GSL with $\varepsilon = 0.04$ the fraction error
is $\sim 10^{-10}$ and the speed is $\sim1, 700\textrm{ geodesics
sec}^{-1}$.  These results can be compared to the publicly available
integral-based {\tt geokerr} code of \citet{DA09}, whose geodesics are
shown as the (more accurate) solid lines in Figure~\ref{fig:rf_geod}.
If we use {\tt geokerr} to sample each geodesic the same number of times
as \grm\ ($\sim 180$), then on the same machine {\tt geokerr} runs at
$\sim1,000\textrm{ geodesics sec}^{-1}$.  It is possible that other
implementations of an integral-of-motion based geodesic tracker could be
faster.

If only the initial and final states of the photon are required, we find
that {\tt geokerr} computes $\sim 77, 000\textrm{ geodesics sec}^{-1}$.
The adaptive Runge-Kutta Cash-Karp integrator in GSL computes $\sim
34, 500\textrm{ geodesics sec}^{-1}$ with fractional error
$\sim10^{-3}$.

\section{Absorption}\label{ABSORB}

\grm\ treats absorption deterministically.  We begin with the radiative 
transfer equation written in the covariant form
\begin{equation}\label{eqn:grradtrans}
\frac{1}{\sC}
\frac{d}{d\lambda} \left(\frac{I_\nu}{\nu^3}\right) =
\left(\frac{j_\nu}{\nu^2}\right) - (\nu\alpha_{\nu,a})
\left(\frac{I_\nu}{\nu^3}\right).
\end{equation}
\citep[see][]{MM84}.  
Here $I_\nu$ is specific intensity and $\alpha_{\nu,a}$ is the absorption 
coefficient (which is always evaluated in the fluid frame).
The absorption coefficient must be computed by a separate subroutine;
for thermal synchrotron emission we set $\alpha_{\nu,a} = j_\nu/B_\nu$.
$\sC$ is a constant that depends on the units of 
$k^\mu$ (in \grm, electron rest mass), $\nu$ (Hertz), and the length 
unit $L$ for the simulation in cgs units.  For \grm\ 
\begin{equation}
\sC = \frac{L h}{m_e c^2}.
\end{equation}
Each quantity in parentheses in equation~\eqref{eqn:grradtrans} is invariant;
$I_\nu/\nu^3$, for example, is proportional to the photon phase space
density.  

Since $I_\nu/\nu^3$ is proportional to the number of photons moving
along each ray, $I_\nu/\nu^3 \propto w$, and equation~\eqref{eqn:grradtrans} 
implies (ignoring emission)
\begin{equation}
\frac{d w}{d\tau_{a}} = -w
\end{equation}
where
\begin{equation}\label{eqn:dtau_abs}
d\tau_{a} = (\nu \alpha_{\nu,a})\; \sC d\lambda
\end{equation}
is the differential optical depth to absorption and the quantity in parentheses is the ``invariant opacity.''  This
equation we integrate with second order accuracy
\begin{equation}\label{eqn:tau_abs_num}
\tau_{a}=\frac{1}{2}\left( 
(\nu \alpha_{\nu,a})_n + (\nu \alpha_{\nu,a})_{n+1}\right)
\; \sC \Delta\lambda,
\end{equation}
and then set
\begin{equation}
w_{n+1} = w_{n} e^{-\tau_{a}}.
\end{equation}
Since the components of $k^\mu$ are expressed in units of the electron
rest-mass energy, $\nu = -k^\mu u_\mu m_e c^2/h$.  Storing the invariant opacity at the end of each step saves computations since it can be reused as the beginning of the following step.

\section{Scattering}\label{SCATTER}

Our treatment of scattering consists of two parts: the first determines
where a superphoton should scatter and the second determines the energy
and direction of the scattered superphoton.

\subsection{Selection of scattering optical depth}

When a superphoton is created or scattered \grm\ selects the scattering
optical depth $\tau_s$ at which the next scattering event will take place.
Scattering follows the cumulative probability distribution 
\begin{equation}\label{eqn:pscatt}
p = 1 - e^{-\tau_{s}}  = \tau_{s} + O(\tau_{s}^2),
\end{equation}
so superphotons will experience on average $\tau_s$ scattering events
when $\tau_s \lesssim 1$.  In optically thin sources this would result
in poor signal to noise in portions of the spectrum dominated by 
scattered light.  To overcome this, we use the biased probability 
distribution
\begin{equation}
p = 1- e^{-b\tau_s}
\end{equation} 
where $b$ is a bias parameter.  This technique was originally proposed by \citet{Kahn50} in the context of deep penetration of neutrons in radiation shielding and has since been extensively explored in the nuclear engineering literature (often refered to as exponential biasing or exponential transform).  Whereas in deep penetration problems $b\le1$ in order to allow for sampling of radiation at high optical depths, here we set $b\ge1$ in order to better sample scattered photons at low optical depths.  Superphotons now experience on average $b\tau_s$ scattering events.  Two superphotons emerge from a scattering
event: the incident superphoton of weight $w$ and a new scattered
superphoton.  For conservative scattering the incident superphoton has
its weight reset to $w (1 - 1/b)$ and the new superphoton has weight
$w/b$, so that weight (photon number) is conserved.

What should we choose for the bias parameter $b$?  The goal is to set $b$ such that scattering is more likely to occur in regions which contribute most to the spectrum.  This is an example of the more general technique of importance sampling.  Typically we set the bias parameter $b = \mathrm{MAX}(1, \alpha
\Theta_e^2 / \tau_{s,max})$ ($\alpha$ is a scaling factor we set to $1 /
\< \Theta_e \>^2$ where $\<\Theta_e\>$ is the volume
averaged dimensionless temperature and $\tau_{s,max}$ is an estimated
maximum scattering optical depth) to improve sampling on the high energy
side of each scattering order, which is populated by photons scattered
from high temperature plasma.  If the bias factor is too large a ``chain
reaction'' results in an exponentially growing number of superphotons;
$1/\tau_{s,max}$ is an estimate of the critical bias factor in a
$\Theta_e = 1$ plasma.  

We evaluate the scattering optical depth along geodesics in a manner
analogous to the absorption optical depth;  
\begin{equation}\label{eqn:deftaus}
\tau_{s}=\frac{1}{2}\left( 
(\nu \alpha_{\nu,s})_n + (\nu \alpha_{\nu,s})_{n+1}\right)
\; \sC \Delta\lambda,
\end{equation}
is the scattering depth along a step.

\subsubsection{Covariant evaluation of extinction coefficient}

In our applications electron scattering dominates.  The general, invariant
expression for the rate of binary interactions $dN_{ab}$ between a
population of particles $dN_a$ and $dN_b$ is
\begin{equation}
\frac{1}{\gdet}\,\frac{dN_{ab}}{d^3x dt} =
\frac{1}{1 + \delta_{ab}}\;
\int \; 
\frac{d^3p_a}{\gdet\,p_a^t} \;
\frac{d^3p_b}{\gdet\,p_b^t} \;
\frac{dN_a}{d^3x \, d^3p_a} \;
\frac{dN_b}{d^3x \, d^3p_b} \;
(-p_{a\mu} p_b^\mu) \; \sigma v_{ab}
\end{equation}
where $\delta_{ab}$ prevents double-counting if $a=b$, $d^3p = dp_1 dp_2
dp_3$, $\sigma$ is the invariant cross section, and $v_{ab} = c (1 + m_a^2
m_b^2/(-p_{a\mu} p_b^\mu)^2)^{1/2}$. This is the manifestly covariant
generalization of equation~(12.7) of \cite{LL75}.  

We want to use this expression to find the cross section for a photon
with wavevector $k_0^\mu$ interacting with a population of particles of
mass $m > 0$.  We therefore set $dN_\gamma/d^3x d^3p = \delta(k^\mu -
k_0^\mu)$ and, dropping the subscripts on $k$ and $p$, the integral
reduces to 
\begin{equation}
\frac{1}{\gdet}\,\frac{dN_{m\gamma}}{d^3x dt} =
\int \; 
\frac{d^3p}{\gdet\,p^t} \;
\frac{dn_m}{d^3p} \;
\frac{(-k_{\mu} p^\mu)}{k^t} \; \sigma c
\end{equation}
where $dn_m = dN_m/d^3x$.  In Minkowski coordinates ($\gdet = 1$),
define $\beta_m \equiv$ the particle speed in the plasma frame and
$\mu_m \equiv$ the cosine of the angle between the particle momentum and
photon momentum in the plasma frame.  Then
\begin{equation}
\frac{dn_{m\gamma}}{dt} = \int \; d^3p \; \frac{dn_m}{d^3p} \; 
(1 - \mu_m \beta_m) \; \sigma c.
\end{equation}
It is convenient to rewrite this rate in terms of a ``hot cross section'' 
\begin{equation}\label{eqn:hotcross}
\sigma_h \equiv {1\over{n_m}} 
\int \; d^3p \; \frac{dn_m}{d^3p} \; 
(1 - \mu_m \beta_m) \; \sigma
\end{equation}
so that the interaction rate for a single photon is $n_m \sigma_h c$ and
the extinction coefficient is 
\begin{equation}
\alpha_{\nu} = n_m \sigma_h.
\end{equation}
So far we have assumed nothing about the interaction process.

\subsubsection{Electron scattering}

For electron scattering the cross section is the Klein-Nishina total
cross section expressed in terms of the photon energy in the electron
rest frame $\equiv \eps_e = \eps \gamma_e (1 - \mu_e \beta_e)$,
($\gamma_e \equiv (1 - \beta_e^2)^{-1/2}$ and we have substituted the
subscript $e$ for $m$):
\begin{equation}\label{eqn:totKN}
\sigma_{KN}=\sigma_{T} \frac{3}{4 \eps_e^2} \left(2 + 
\frac{\eps_e^2 (1 + \eps_e)}{ (1 + 2\eps_e)^2} + 
\frac{\eps_e^2 - 2\eps_e - 2 }{2 \eps_e}\log(1+ 2\eps_e)\right).
\end{equation}
Here $\sigma_T$ is the Thomson cross section.  For $\eps \ll 1$, 
\begin{equation}
\sigma_{KN} = \sigma_{T} (1 - 2 \eps + O(\eps^2))
\label{eqn:totT}
\end{equation} 
which is numerically stable for small $\eps$, unlike equation~\eqref{eqn:totKN}.

Typically we assume a thermal electron distribution,
\begin{equation}
\frac{dn_e}{d\gamma_e} = \frac{n_e}{\Theta_e}\frac{\gamma_e^2
\beta_e}{K_2(\Theta_e^{-1})} \exp\left(-\frac{\gamma}{\Theta_e}\right),
\end{equation}
and evaluate (\ref{eqn:hotcross}) by direct integration to obtain
$\sigma_h(\Theta_e,\eps)$.  It is efficient to store the resulting cross
sections in a two-dimensional lookup table at the beginning of the
calculation.  Our $\sigma_h$ agrees with Wienke (1985).

\subsection{Scattering kernel}

Once it is determined that a superphoton should be scattered at an event
$x^\mu_s$, the superphoton is passed to a scattering kernel which
processes the scattering event according to the following procedure.
Only unpolarized light is considered.

First, a plasma frame orthonormal tetrad is constructed by the same
Gram-Schmidt orthogonalization procedure described in \S\ref{MANUFACTURE},
and the old superphoton wave vector is transformed from the coordinate
frame to the tetrad frame.

Second, a scattering electron is selected.  We use the procedure
described by \citet{C87}, which selects the four-momentum $p^\mu_e$ of
the scattering electron with an efficiency of $72\%$ at $\Theta_e = 1$
and nearly $100\%$ for $\Theta_e\ll1$ or $\Theta_e\gg1$ \citep{C87}.

Third, we boost from the plasma frame tetrad to an electron frame tetrad
$q_{(a)}^\mu$ and construct the scattered photon wave vector in this
frame.  The differential scattering cross section is sampled for the
scattered photon energy $\eps_e'$ and the scattering angle $\theta$.
For low energy photons ($\eps_e < \eps_l$; in \grm\, $\eps_l = 10^{-4}$)
the scattering is approximately elastic, so we set $\eps_e' = \eps_e$
and sample the Thomson differential cross section
\begin{equation}
\frac{2\pi}{\sigma_T}\frac{d\sigma_T}{d\cos\theta} = \frac{3}{8} (1 +
\cos^2\theta)
\end{equation}
for the scattering angle $\theta$ using a rejection scheme.  For $\eps_e >
\eps_l$ we sample the Klein-Nishina differential cross section 
\begin{equation}
\frac{2\pi}{\sigma_T} {{d\sigma_{KN}}\over{d\eps_e'}} =
{1\over{\eps_e^2}}({\eps_e\over{\eps_e'}} + {\eps_e'\over{\eps_e}} -
1 + \cos^2\theta),
\end{equation}
for $\eps_e'$ using a rejection scheme.  Here $\cos\theta = 1 + 
1/\eps_e - 1/\eps_e'$.  This procedure is inefficient for $\eps_e \gg 1$, 
but in our target application such large photon energies are rare.  
Drawing a final angle $\phi$ from a uniform distribution on $[0,2\pi)$ 
completes the specification of the photon wave vector
\begin{subequations}
\begin{align}
k^{(0)} &= \eps_e' \\
k^{(1)} &= \eps_e'\cos\theta \\
k^{(2)} &= \eps_e'\sin\theta\cos\phi \\
k^{(3)} &= \eps_e'\sin\theta\sin\phi 
\end{align}
\end{subequations}
in the electron-frame tetrad basis $q_{(a)}^\mu$; $q_{(1)}$ is aligned 
parallel to the spatial part of the incoming photon wave vector.

Finally, we boost back from the electron-frame tetrad to the plasma
frame tetrad (some of these steps are combined in our code for
computational efficiency), and use the plasma frame basis vectors to
obtain the coordinate frame scattered photon wave vector $k'_\mu$.

\section{Spectra}\label{CENSUS}

Spectra can be measured using a ``detector'' with area $\Delta A$ at
distance $R$, frequency channels of logarithmic width $\Delta\ln\nu$,
and integration times $\Delta T$.  The flux density in frequency bin 
$i$ is then just
\begin{equation}\label{FLUXEQ}
F_{\nu,i} = \frac{1}{\nu_i \Delta\ln\nu\; \Delta T\; \Delta A} 
\sum_j w_j (h \nu)_j
\end{equation}
where the sum is taken over all superphotons $j$ that land in the
channel during the integration.  In principle a software detector can
behave just like a physical detector, producing time-dependent spectra
from time-dependent flow models.  In practice time-dependent models are
not (yet) treated self-consistently.

To see why, consider a time-dependent model based on a general
relativistic magnetohydrodynamics (GRMHD) model of a black hole
accretion flow.  Self-consistent treatment of the radiation field would
require generating and tracking superphotons through the simulation data
as it evolves, i.e. coupling the \grm\ to the GRMHD code (the simulation
data could be stored and post-processed, but this would require storing
almost every timestep and would be impractical and inefficient).  The
mean number of superphotons tracked simultaneously would depend on the
desired signal-to-noise in the final spectrum, as well as the time and
energy resolution.  Our experience suggests that for nominal energy
resolution, signal-to-noise, and time resolution of order the horizon
light crossing time, $\sim 10^8$ superphotons would need to be maintained within the simulation for the duration of the evolution.  The total number required is then $N_{tot}\gg10^8$ and is currently inaccessible without significant computational resources.
We plan to attempt this calculation later.

For now we construct spectra of time-dependent data using a
stationary-data (or ``fast-light'') approximation: each time-slice of
data is treated as if it were stationary (time-independent).  The slice
emits superphotons for a time $\Delta t$.  The photons then propagate
through the slice data (as $t$ varies along a geodesic the fluid
variables are held fixed) and are detected at large distance.  The
time-steady flux is obtained by substituting $\Delta t$ for $\Delta T$
in equation~\eqref{FLUXEQ}.  

\subsection{Measuring $\nu L_\nu$}

In practice we measured $\nu L_{\nu,i}$ rather than $F_{\nu,i}$.  Since $\nu
L_\nu = 4 \pi R^2 \nu F_\nu$, and $R^2/\Delta A$ is the solid angle
$\Delta\Omega$ occupied by the detector, 
\begin{equation}
\nu L_{\nu,i} = {4\pi\over{\Delta\Omega\Delta t}} {1\over{\Delta\ln\nu}} \; 
\sum_j w_j h \nu_j.
\end{equation}
Typically our ``detectors'' capture all the superphotons in a large 
angular bin $\Delta\Omega$ around the source.  For example, in studies
of axisymmetric black hole accretion flows the angular bins capture all
photons with Boyer-Lindquist $r > 100 G M/c^2$, $\theta_{n} < \theta <
\theta_{n+1}$, independent of $\phi$, where $\theta_n$ are the bin
boundaries.

We can also estimate average values for any quantity $Q$ associated with
emission from a source (e.g. the absorption optical depth).  We define
the weight-averaged value of $Q$ via
\begin{equation} 
\< Q \> \equiv {{\sum Q w}\over{\sum w}},
\end{equation}
where the sum is taken within an energy and angular bin.

\subsection{Optimal weights}\label{WEIGHTS}

The fractional variance in the Monte Carlo estimate for $\nu L_\nu$ is
proportional to $\overline{w^2}/(\overline{N_s} \overline{w}^2)$, where
overline means expectation value and $N_s$ is the number of superphotons
in the bin.  Evidently the optimal weighting of superphotons is achieved
when (1) the weights of superphotons are the same within bins, and (2)
the superphotons are evenly distributed across frequency bins.

If, as we created new superphotons, we knew $\nu L_\nu$, then we could
set
\begin{equation}
w_\nu = \frac{1}{\overline{N_s} h} \; L_\nu \Delta \ln\nu
\end{equation}
and set $\overline{N_s} = N_{s,tot}/N_b$, where $N_b$ is the number of 
frequency bins.  

Of course we do not know $L_\nu$, but for the special case of an optically
thin emitting plasma we can estimate it:
\begin{equation}
L_{\nu} \approx \int d^3 x \sqrt{-g} \int d\Omega \; j_{\nu}
\end{equation}
This estimate assumes that all photons escape to infinity, and it
ignores Doppler shift, gravitational redshift, scattering, and angular
structure in $\nu L_\nu$.  Nevertheless it is useful because (1) it can
be calculated before the Monte Carlo calculation begins; (2) it is far
better to use the information contained in this rough estimate of the
spectrum than to proceed using, e.g., uniform weights.  

\section{Tests}\label{TESTS}

We verify the accuracy of \grm\ by comparing spectra produced on
idealized problems against a reference spectrum $(\nu L_\nu)_{ref}$
computed analytically (when possible) or computed by an independent
code.  For all tests we use the following error norm, which effectively
measures the maximum of the fractional error, compared to the reference
solution, over frequency:
\begin{equation}
\eno = \frac{1}{\Delta\ln\nu} \int_{\nu_{min}}^{\nu_{max}} 
\frac{|(\nu L_\nu)_{\grm} - (\nu L_\nu)_{ref}|}{(\nu L_\nu)_{ref}} 
d\ln\nu
\end{equation} 
where $\Delta\ln\nu=\ln(\nu_{max}/\nu_{min})$ is the range of
integration.  The range of integration is the same as plotted in
the spectra for each test.

\subsection{Optically thin synchrotron sphere}

\begin{figure}
\plotone{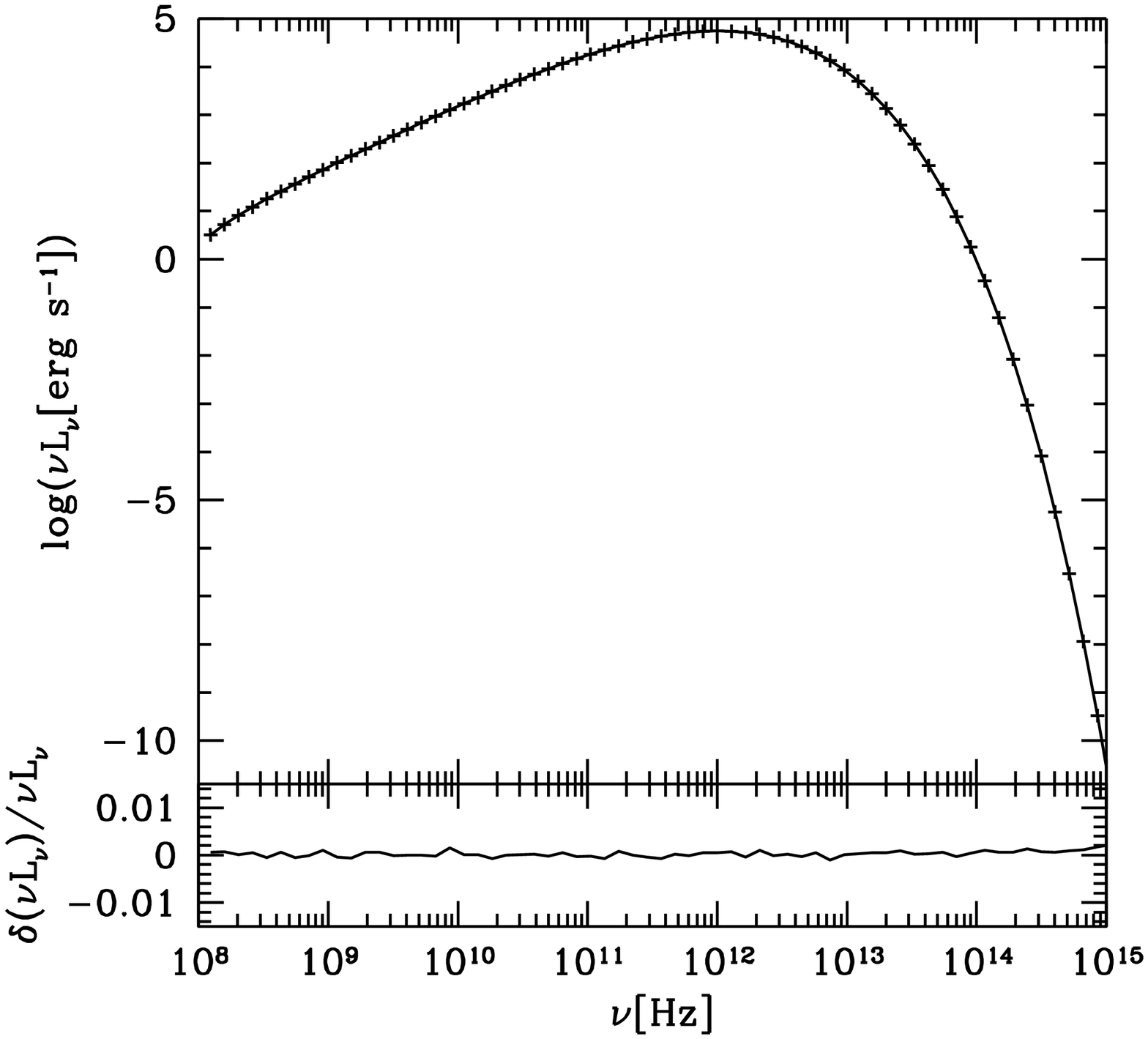}
\caption{In the top panel, the spectrum of a
synchrotron emitting sphere with low optical depth above $\sim10^8\textrm{ Hz}$ 
viewed nearly perpendicular to the magnetic field 
from \grm\ (crosses) and from a semi-analytic procedure (solid line).
The bottom panel shows the fractional difference between the two results.}
\label{fig:spec_athin}
\end{figure}

\begin{figure}
\plotone{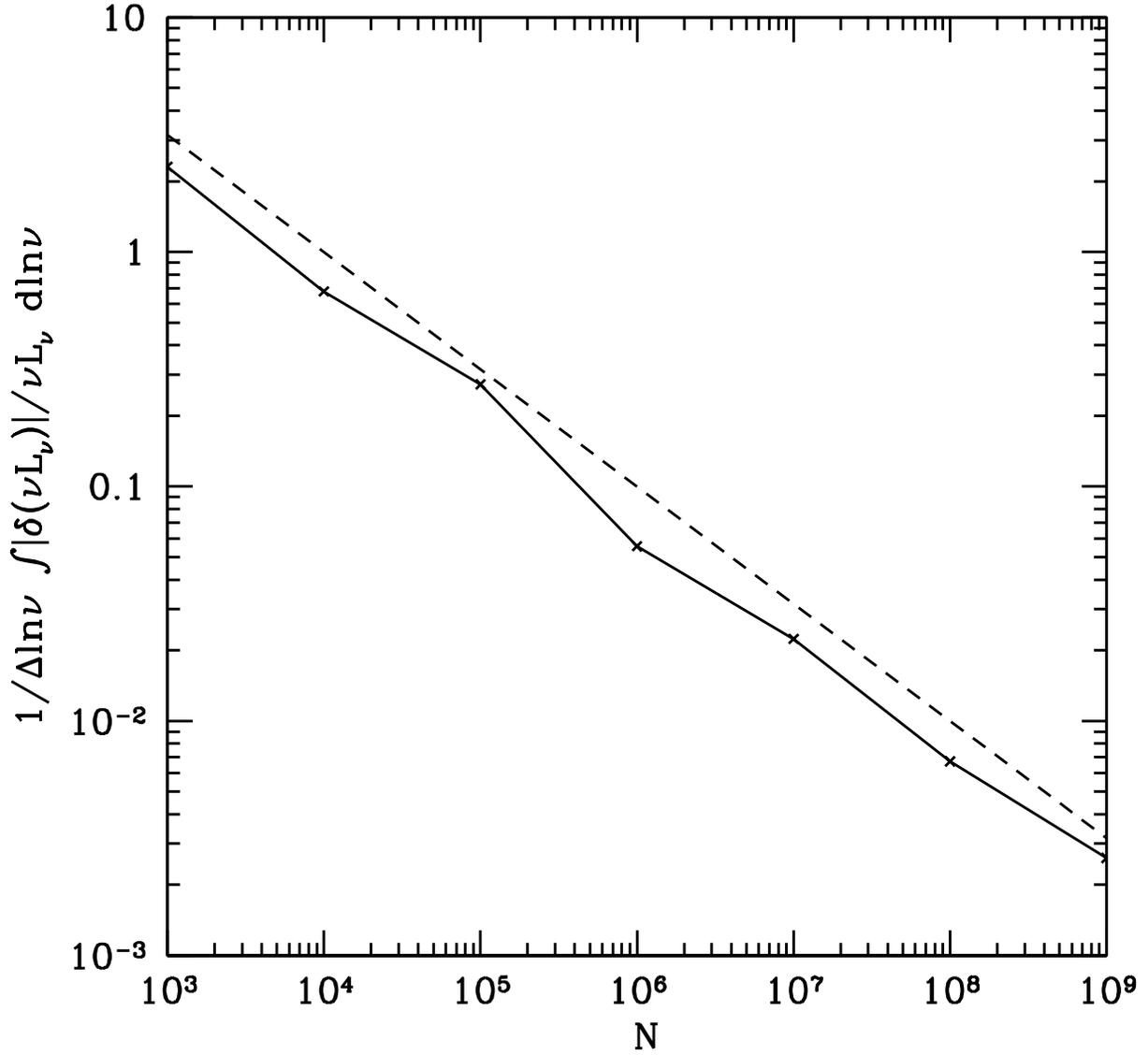}
\caption{Integrated fractional error in the \grm\ spectrum for a
synchrotron emitting sphere with low optical depth 
viewed nearly perpendicular to the magnetic field as a function of the 
number of superphotons produced.  The results are similar for other 
magnetic field orientations.  The dashed line is proportional to $N^{-1/2}$.}
\label{fig:l1_athin}
\end{figure}

First we consider emission from a homogeneous, optically thin spherical
cloud of unit volume threaded by a vertical magnetic field in flat
space.  The cloud parameters are $\Theta_e=100$, $B=1\textrm{ G}$, and
$n_e=10^{15}\textrm{ cm}^{-3}$, which gives an optical depth at
$\nu=10^9\textrm{ Hz}$ of $\sim10^{-2}$ perpendicular to the magnetic
field. The emissivity and absorptivity are constant along any line of
sight, so 
\begin{equation}
I_{\nu} = \frac{j_{\nu}}{\alpha_{\nu}}(1 - e^{-\alpha_{\nu}L})
\approx j_{\nu} L + O(\tau_a^2)
\end{equation}
where $L$ is the path length through the sphere.  We numerically
integrate this expression over detector solid angle to compute the
spectral energy distribution, and compare with the spectrum \grm\
produces in Figure~\ref{fig:spec_athin}.  Evidently the result is
unbiased.   Figure~\ref{fig:l1_athin} demonstrates that \grm\ converges on
the correct solution as $\propto N^{-1/2}$.

\subsection{Optically thick synchrotron sphere}

The second test is identical except that the electron number density and
therefore the optical depth are increased by a factor of $10^5$.  The
intensity along any line of sight is again
\begin{equation}
I_{\nu} = \frac{j_{\nu}}{\alpha_{\nu}}(1 - e^{-\alpha_{\nu}L})
\end{equation}
The emitting region becomes optically thick when $(j_{\nu}/B_{\nu}) L
\gtrsim 1$.  For the thermal synchrotron emissivity we use, this occurs
below a critical frequency.  

\begin{figure}
\plotone{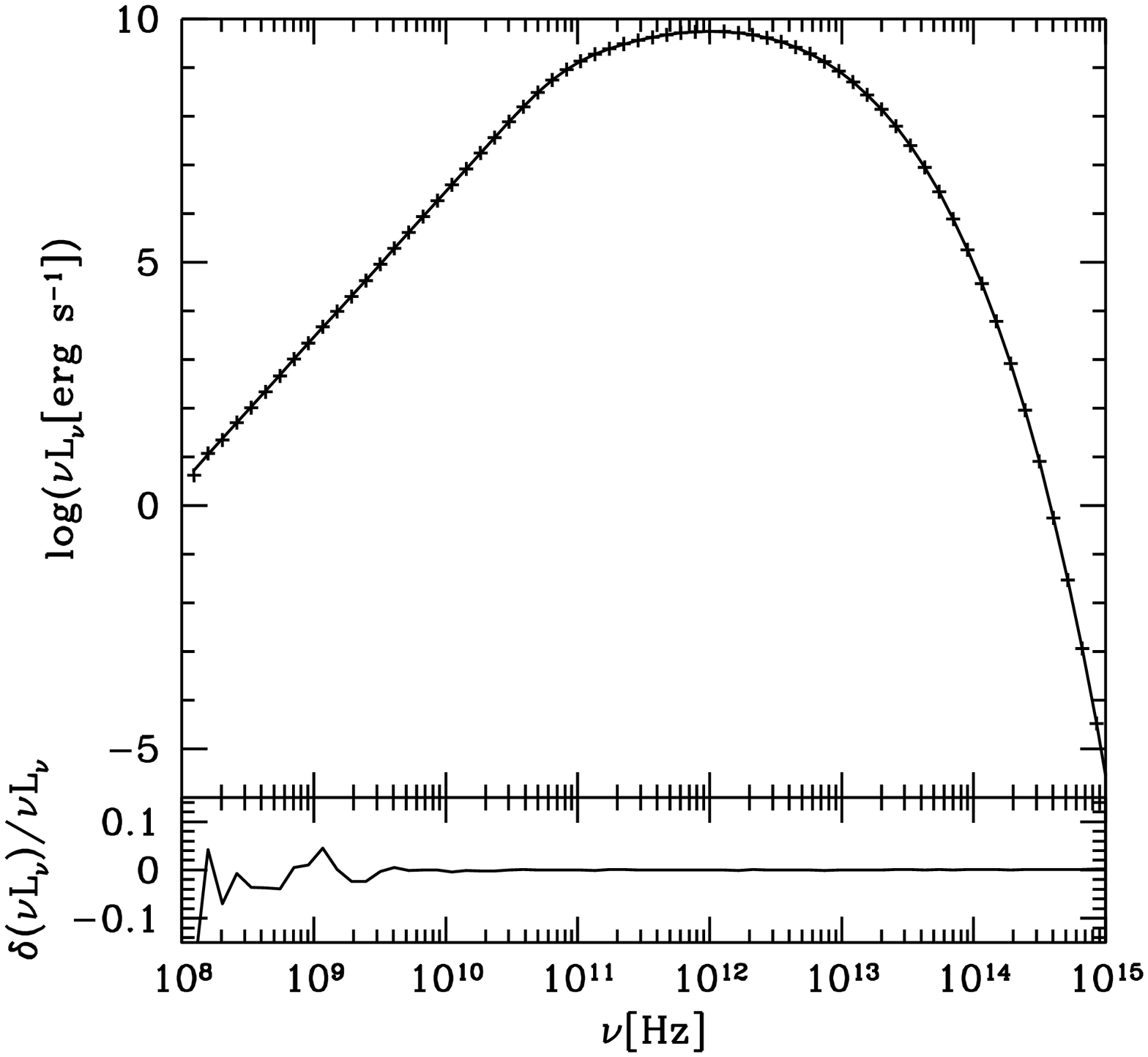}
\caption{In the top panel, the spectrum of a synchrotron emitting sphere
of high optical depth below $\sim10^{11}\textrm{ Hz}$ viewed nearly perpendicular to the magnetic field 
from \grm\ (crosses) and for a semi-analytic procedure (solid line).
The bottom panel shows the fractional 
difference between the two results.}
\label{fig:spec_athick}
\end{figure}

\begin{figure}
\plotone{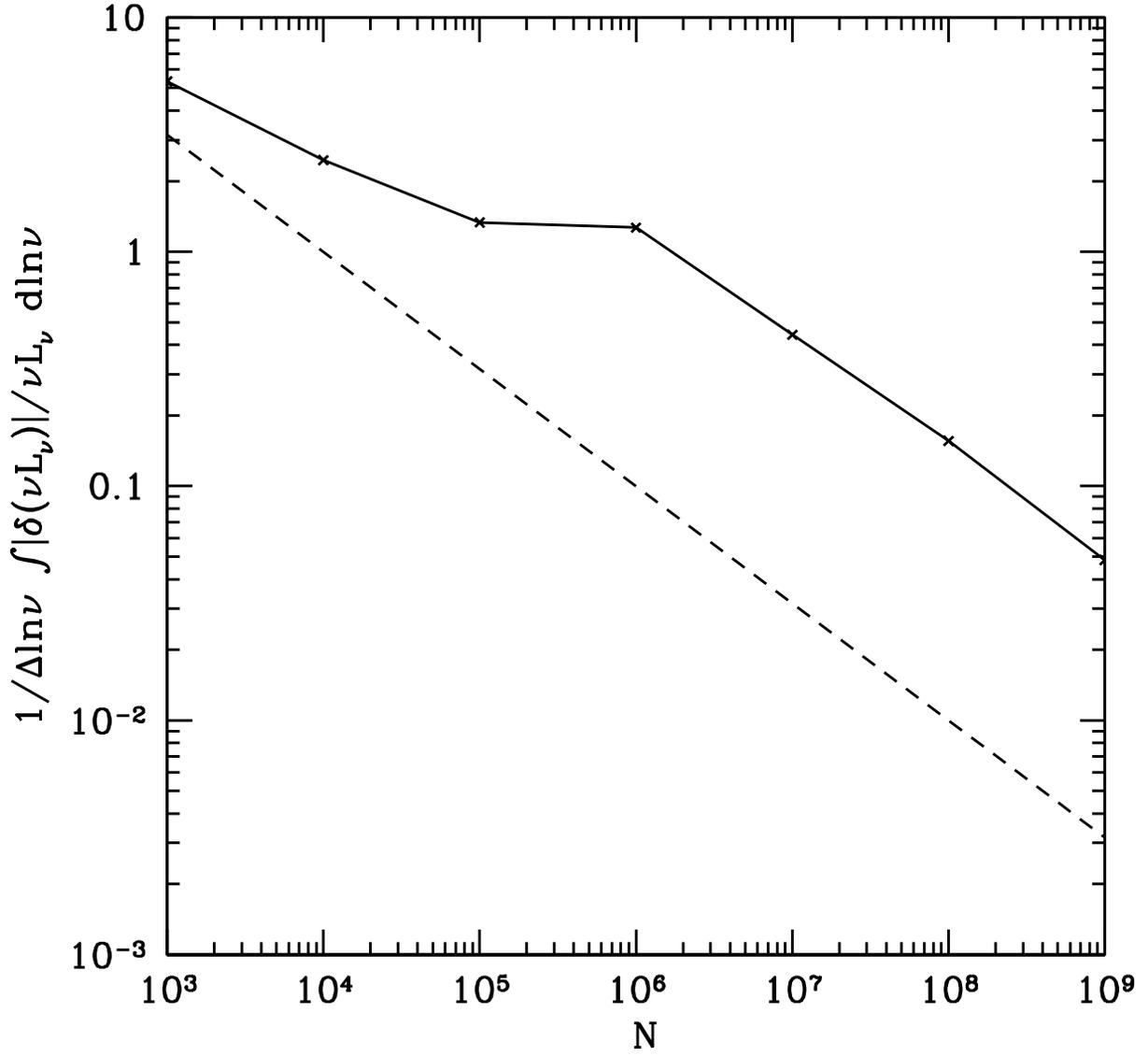}
\caption{Integrated fractional error in the \grm\ spectrum of a
synchrotron emitting sphere with high optical depth 
as a function of the number of superphotons produced.  The 
dashed line is proportional to $N^{-1/2}$.}
\label{fig:l1_athick}
\end{figure}

The spectrum is shown in Figure~\ref{fig:spec_athick} and the convergence
is shown in Figure~\ref{fig:l1_athick}.  The figures make two key points:
convergence is slow for small numbers of superphotons; and the overall
magnitude of the error is larger than in the optically thin case shown
in Figure~\ref{fig:l1_athin}.  

The slow initial convergence is due to the large optical depth at some
frequencies.  When the optical depth is large no superphotons of
appreciable weight are recorded until some superphotons have been
created in the fraction $\sim 1/\tau$ of the volume that lies within the
photosphere.  Our problem has $\tau\sim10^5$ at $\nu\sim10^8 \textrm{
Hz}$.  Since $\eno$ effectively measures the maximum of the error
over frequency, it is not surprising that \grm\ requires $\sim10^6$
superphotons before it begins to converge as $N^{-1/2}$.

\subsection{Comptonization of soft photons in a spherical cloud of plasma}

This test is based on a problem posed in \S 6 of \citet{PSS}: the
spectrum of a spherical, homogeneous, unmagnetized cloud of radius $R$
that contains thermal electrons at density $n_e$ and temperature $T_e$
that scatter light from a central, thermal source of temperature $T_s$.
Absorption and emission in the cloud are neglected.  The dimensionless
parameters of the problem are $\Theta_e= k T_e/(m_ec^2)$, $\tau =
R\sigma_T n_e$, and $\Theta_s = k T_s/(m_e c^2)$.

For this test our reference spectrum is computed with an implementation
of Pozdnyakov et al.'s Monte Carlo scheme kindly provided by S. Davis.
This code, {\tt sphere}, has been modified in two ways: we have replaced
the approximate hot cross sections defined in \citet{PSS} with our more
exact, numerically integrated values, and we use the exact Klein-Nishina
cross section equation~\eqref{eqn:totKN} when choosing the electron with
which a photon should scatter.  Without these changes to {\tt sphere}
differences between the spectra are $\lesssim1\%$, consistent with the
error \citet{PSS} quote for their approximations.  These small
differences are enough, however, to prevent \grm\ from converging as
expected for large numbers of superphotons. 

\begin{figure}
\plotone{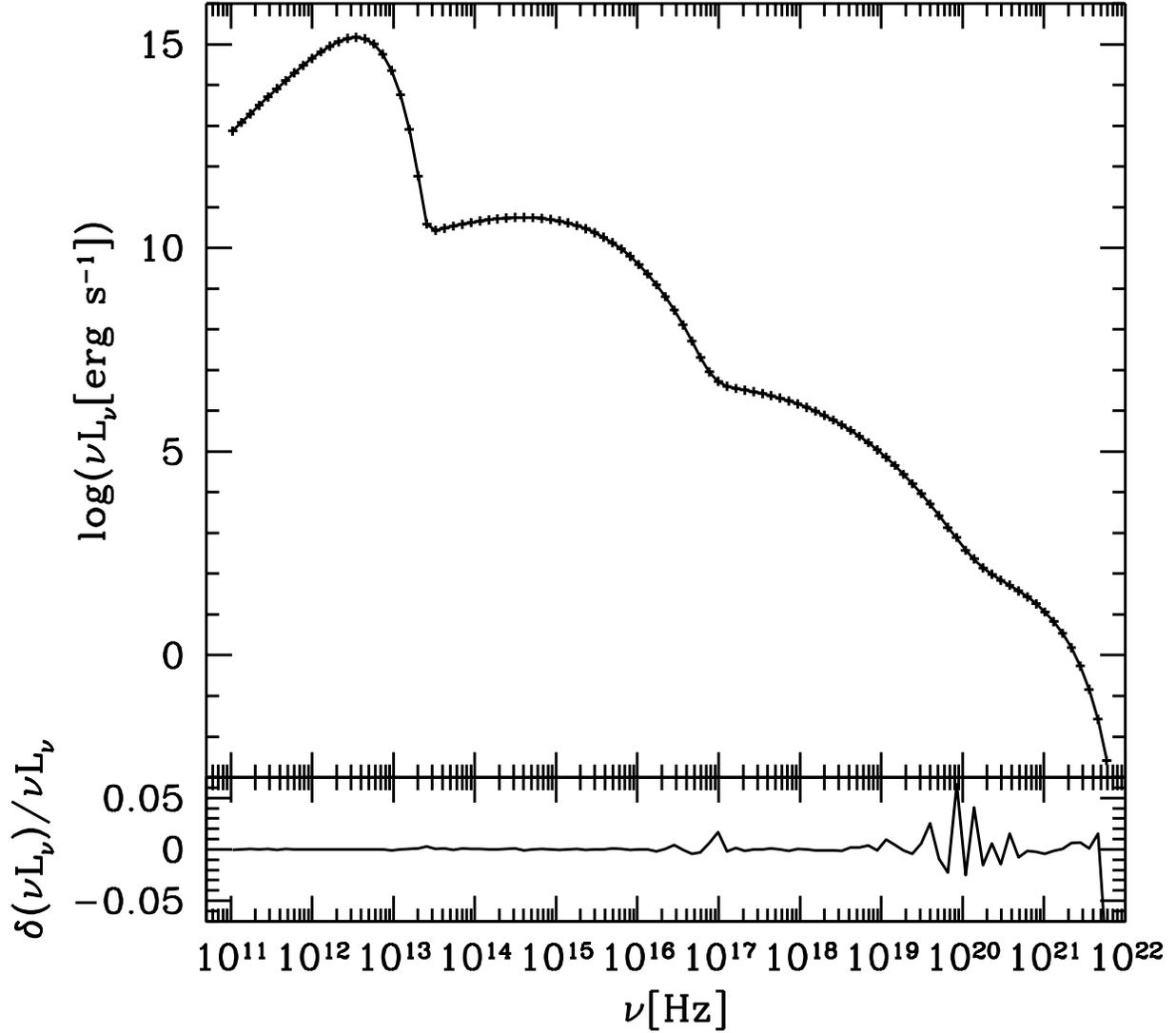}
\caption{Spectra (upper panel) from \grm\ (points)
and {\tt sphere} (solid line) produced by Comptonization of soft photons
in a homogeneous, spherical cloud of hot plasma. 
Computations are done for: plasma optical thickness $\tau=10^{-4}$, 
plasma temperature $\Theta_e$=4 and the central source radiative 
temperature $kT_r/m_ec=10^{-8}$.
Lower panel shows the fractional difference between the two spectra.}
\label{fig:psstest_1}
\end{figure}

\begin{figure}
\plotone{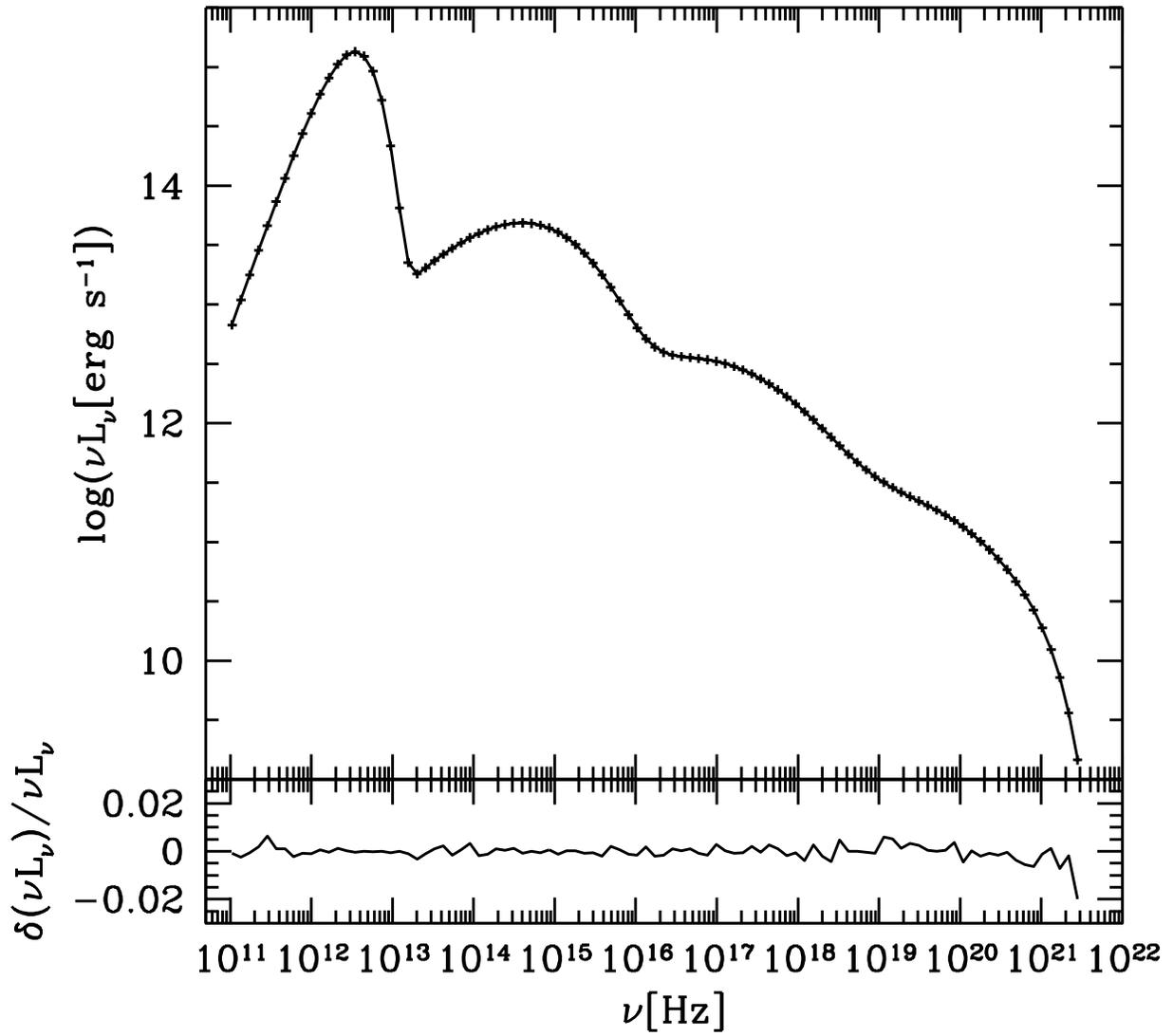}
\caption{Same as in Figure~\ref{fig:psstest_1}, but for $\tau$=0.1.}
\label{fig:psstest_2}
\end{figure}

\begin{figure}
\plotone{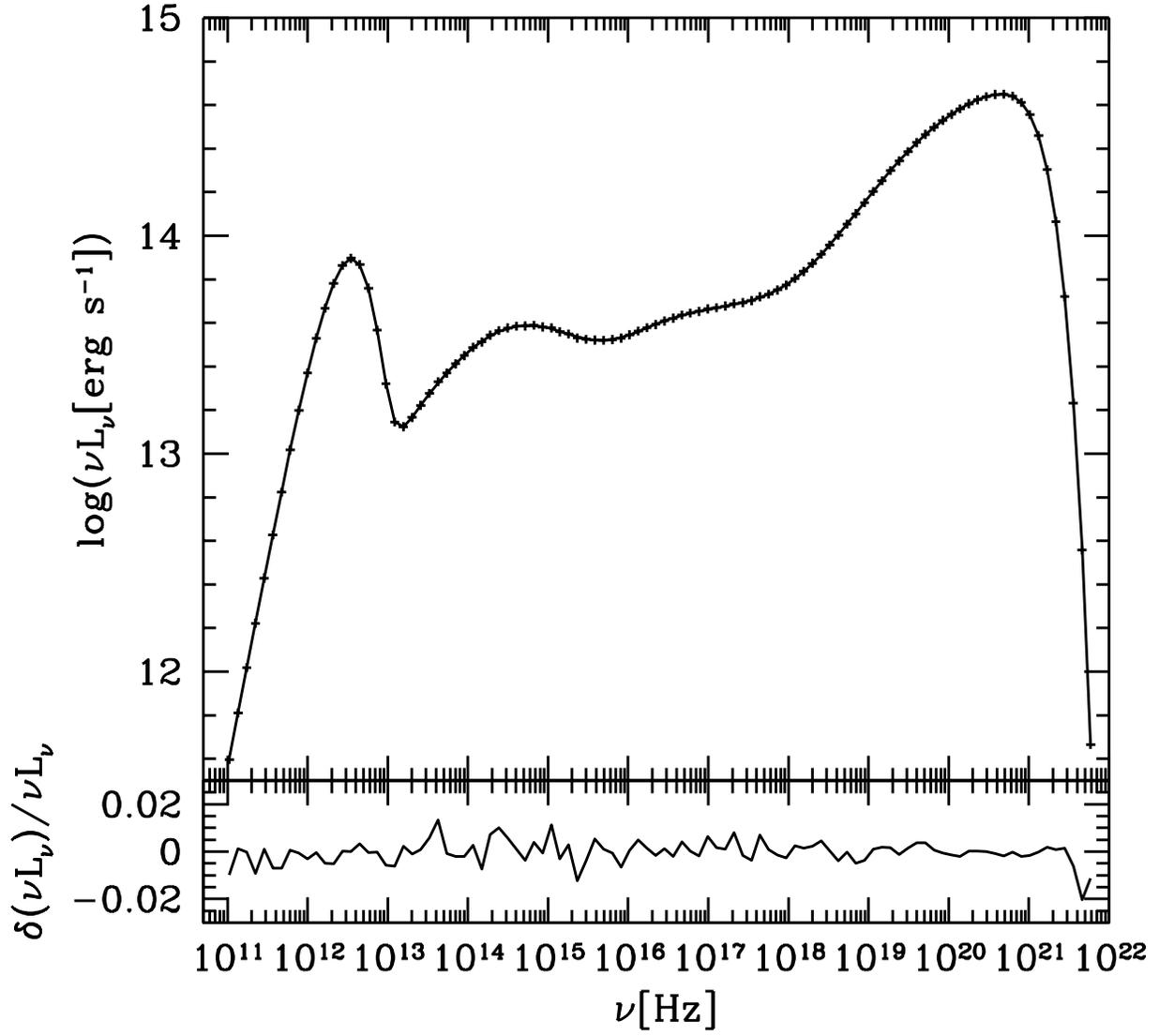}
\caption{Same as in Figure~\ref{fig:psstest_1}, but for $\tau$=3.0.}
\label{fig:psstest_3}
\end{figure}

\begin{figure}
\plotone{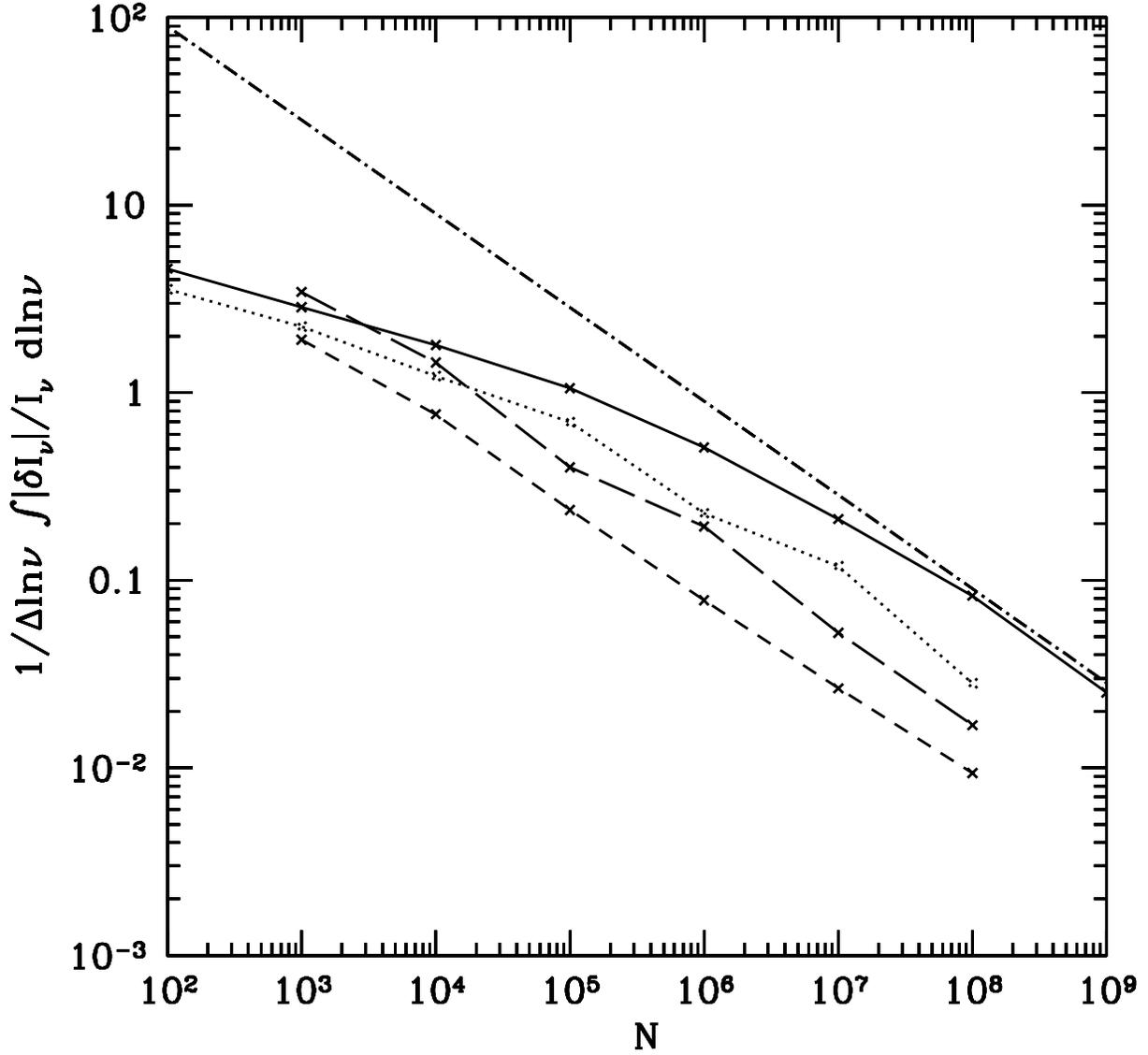}
\caption{Integrated fractional difference between \grm\ and {\tt sphere} for the spherical scattering test for optical depths of $10^{-4}$ (solid), $0.1$ (short dash), and $3$ (long dash).  The dotted line shows the self-convergence results for the {\tt sphere} code for an optical depth $\tau=10^{-4}$.  The dot-dash line is proportional to $N^{-1/2}$.}
\label{fig:l1_pss}
\end{figure}

Figures~\ref{fig:psstest_1},~\ref{fig:psstest_2} and~\ref{fig:psstest_3}
show the radiation spectra (upper panels) produced by both codes and a
fractional difference (bottom panels) between them for $\Theta_e=4$,
$\Theta_s=10^{-8}$ and for various values of the optical depth $\tau$ =
$10^{-4}$, 0.1, and 3.  Figure~\ref{fig:l1_pss} demonstrates that \grm\
converges to the reference solution as $\propto N^{-1/2}$ for each
optical depth.

\subsection{Synchrotron self-absorbed spectra in black hole spacetimes}

We consider two idealized problems: (1) smooth, spherically symmetric
infall onto a Schwarzschild black hole; (2) a snapshot of a turbulent
accretion flow around a Kerr black hole with $a/M=0.9375$ produced by
general relativistic MHD simulation with the {\tt HARM} code
\citep{GMT}.  In this test, our reference solution is computed by the
ray tracing code \ibothros\ \citep{noble07}.  \ibothros\ solves the
invariant form of the radiative transfer equation along geodesics that
terminate in a fictitious camera at large distance.  Spectra are
constructed by imaging the source at successive frequencies and
performing an angular integral over the images to estimate $\nu L_\nu$.
In these tests scattering is turned off in \grm.

There are algorithmic differences between \grm\ and \ibothros\ that lead
to differences in their spectra.  

First, \grm\ measures the flux in energy and angular bins, whereas
\ibothros\ measures the flux at a particular inclination and energy.
This is not an important effect unless there is sharp angular or energy
structure in the spectrum.

\begin{figure}
\plotone{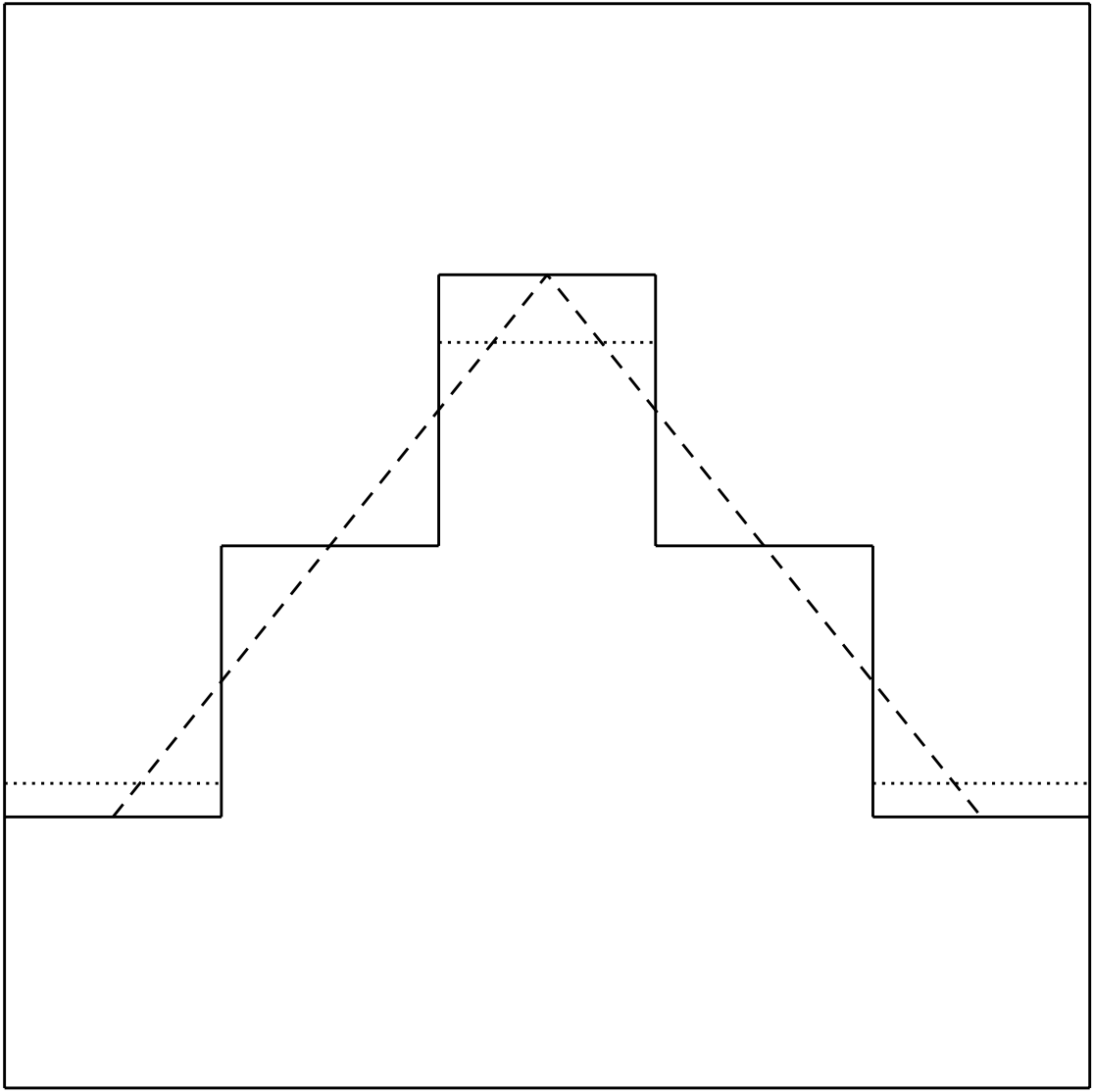}
\caption{Illustration of how interpolation can lead to a discrepancy
between \grm\ and \ibothros\ when the spectrum is 
sensitive to grid-scale structure.  Shown are the grid specified values for some fluid property (solid line), the 
interpolated values (dash line), and the average zone values based on interpolation (dotted line).}
\label{fig:interp_problem}
\end{figure}

The next difference is more subtle and is related to the treatment of
gridded model data used to construct these tests.  In \grm\, quantities
such as the density, temperature, etc., are viewed as the average of
these variables over a grid zone.  In \ibothros\, the grid variables are
viewed as zone-centered samples and a continuous distribution is created
by multi-linear interpolation between zone centers.  The difference is
illustrated in one dimension in Figure~\ref{fig:interp_problem}.
\ibothros\ and \grm\ therefore differ in zone-averaged emissivity by
$O(\Delta x/L)^2$, where $\Delta x$ is the zone size and $L$ is the
characteristic scale of the emitting structure.

Differences in the \grm\ and \ibothros\ spectra of structures with
$\Delta x \ll L$ are therefore small.  High frequency synchrotron
emission, however, is exponentially dominated by emission from a few
zones with highest $\nu_s$ (see equation~\eqref{eqn:nu_nus}; these are the zones with
highest temperature or strongest magnetic field).  Then $\Delta x/L \sim
1$ for high frequency emission, and the \grm\ and \ibothros\ spectra
differ by of order unity.  A similar effect occurs for low frequency
synchrotron emission where the optical depth is large.  The spectrum is
sensitive to the run of physical variables through the photosphere,
which has size comparable to or smaller than a grid zone.  The subsequent test results will omit these parts of the spectra.  These
inconsistencies between \grm\ and \ibothros\ could be eliminated by
using identical, continuous models for subgrid reconstruction.  But this
would require significant investment in recoding that, in our view, is
not worthwhile: to the extent that the spectrum depends on the flow
structure at and below the grid scale, it is not reliable!

In spite of these differences, and other differences between the codes
related to differing accuracy parameters, \grm\ should converge on
the \ibothros\ result until the effects of data interpolation become
the dominant sources of error.

\begin{figure}
\plotone{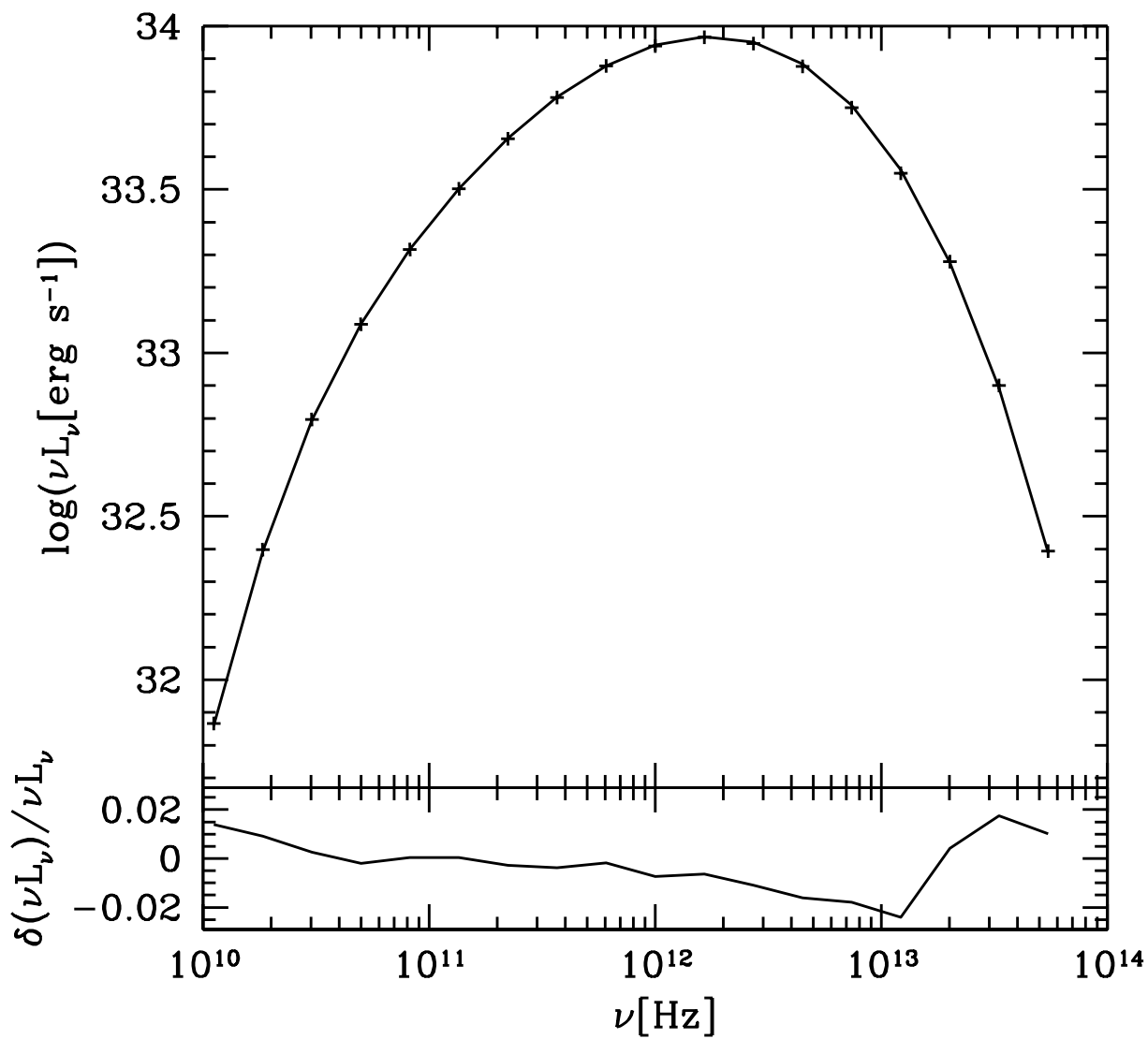}
\caption{The top panel shows the spectrum of a radially infalling spherically symmetric source threaded with a radial magnetic field around 
a Schwarzschild black hole as computed by \ibothros\ (solid line) and
\grm\ (crosses).  The bottom panel shows the fractional difference 
between the two.}
\label{fig:spec_schw}
\end{figure}

\begin{figure}
\plotone{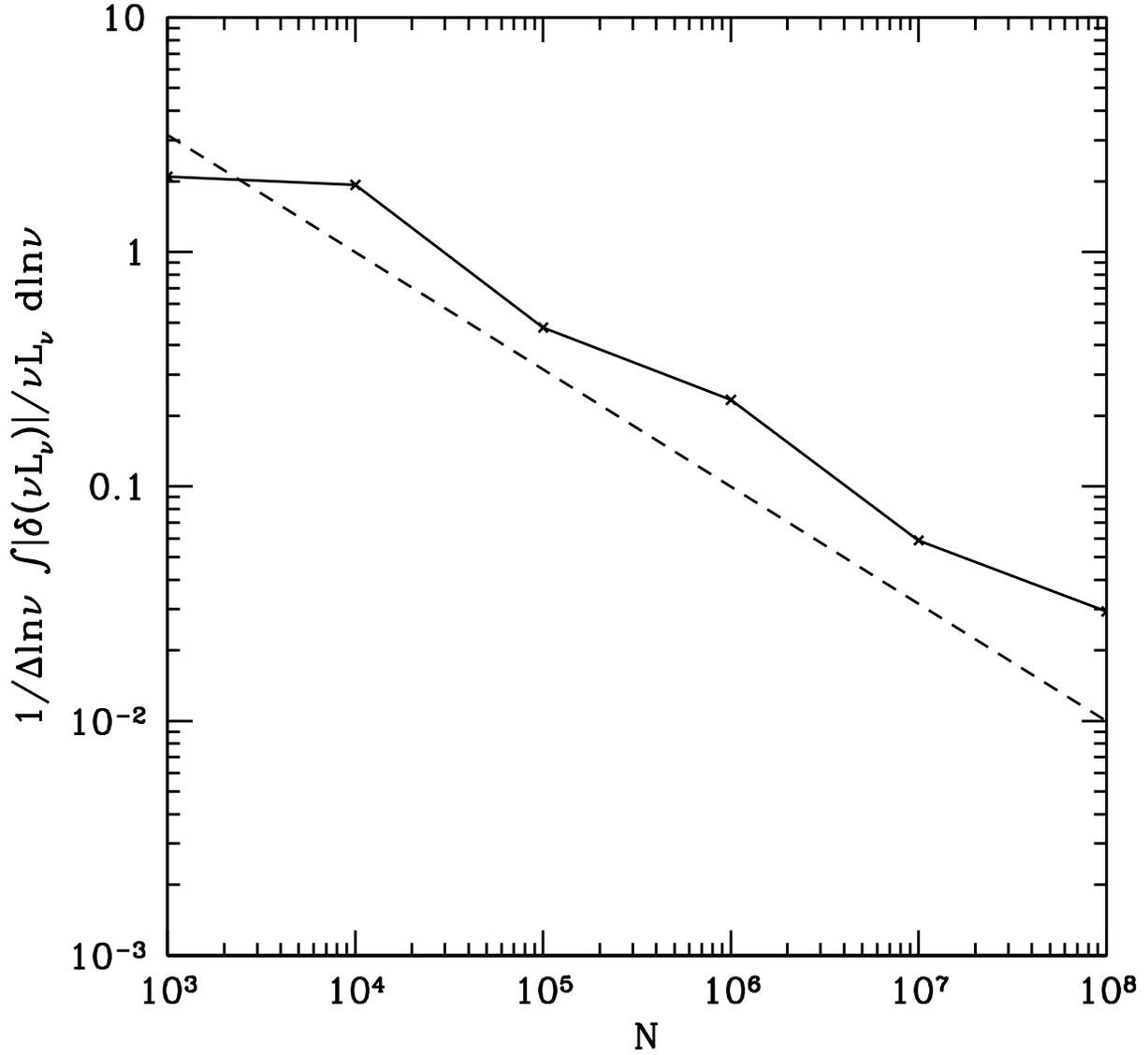}
\caption{Integrated fractional error in the \grm\ spectrum for the spherically 
symmetric Schwarzschild problem as a function of the number of superphotons 
produced.  The dashed line is proportional to $N^{-1/2}$.}
\label{fig:l1_schw}
\end{figure}

Figures~\ref{fig:spec_schw}~and~\ref{fig:l1_schw} show the spectrum and
convergence relative to \ibothros, respectively, for the spherical
accretion problem.  This problem is an attractive test for at least two
reasons.  First, the emission is isotropic so the effects of angular
binning in \grm\ are eliminated.  Second, the flow is smooth, so the
differences due to data interpolation will be small.  Evidently the
spherical accretion model converges at the expected, $N^{-1/2}$, rate.

\begin{figure}
\plotone{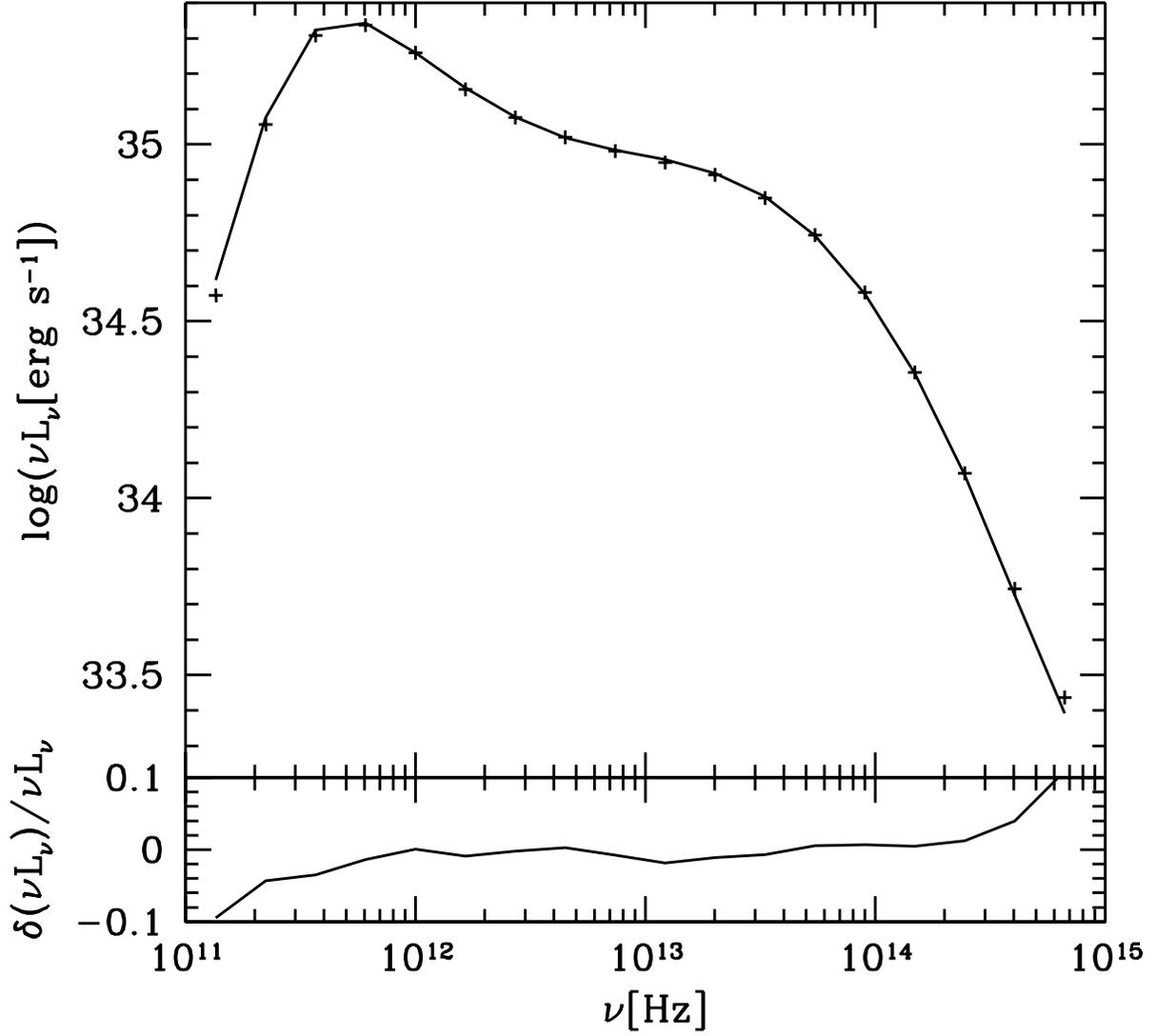}
\caption{The top panel shows the spectrum of a snapshot from a {\tt HARM} simulation of a turbulent accretion flow onto a Kerr black hole 
as computed by \ibothros\ (solid line) and \grm\ (crosses).  The bottom panel shows the fractional difference 
between the two.}
\label{fig:spec_harm2}
\end{figure}

\begin{figure}
\plotone{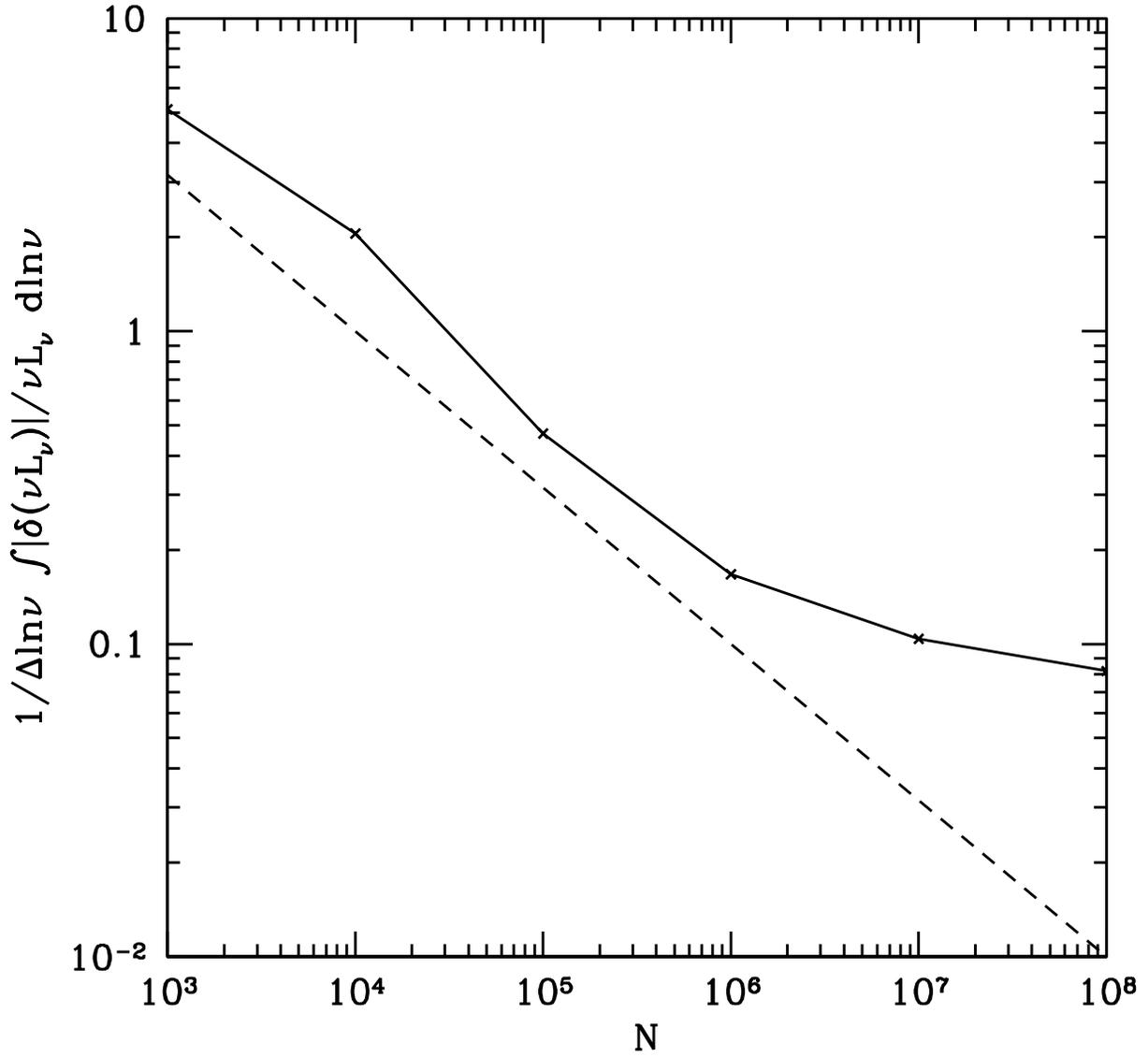}
\caption{Integrated fractional error in the \grm\ spectrum for the turbulent accretion problem 
as a function of the number of superphotons produced.  The dashed line is proportional to $N^{-1/2}$.}
\label{fig:l1_harm2}
\end{figure}

Figures~\ref{fig:spec_harm2}~and~\ref{fig:l1_harm2} show the spectrum and
convergence relative to \ibothros, respectively, for the turbulent
accretion problem.  This comparison is more challenging because the high
energy emission originates in compact ``hot spots.''  Evidently the two
models agree at the $10\%$ level everywhere, with the largest differences at high and
low frequency, where the subgrid reconstruction comes into play.  Excluding
these high and low frequency regions, the agreement between the codes is
at the few percent level.

\section{Sample Calculation \& Full Code Tests}\label{SAMPLE}

\begin{figure}
\plotone{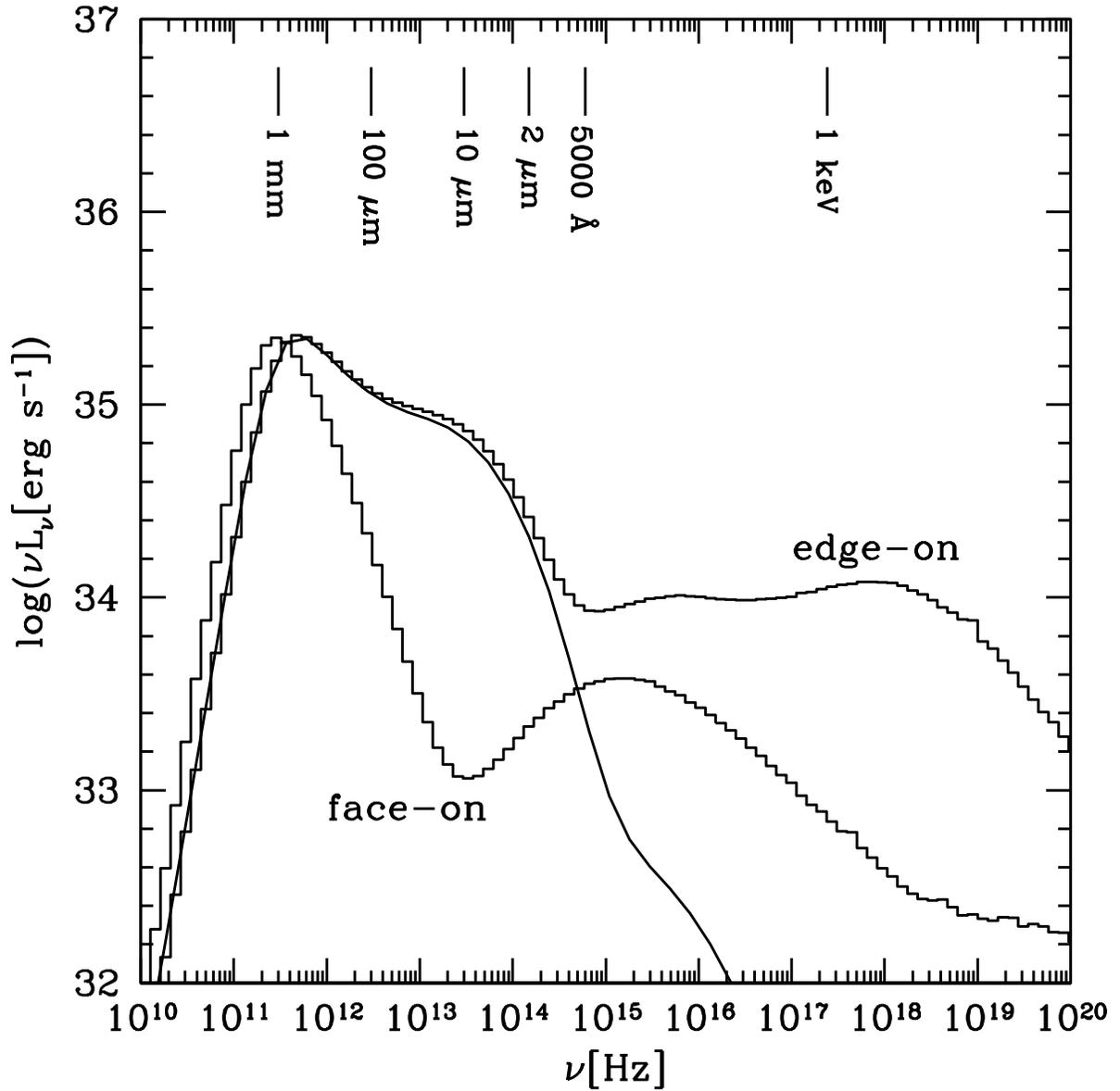}
\caption{Same as Figure~\ref{fig:spec_harm2} except Compton scattering is included.  The histograms show the \grm\ result for nearly edge-on and face-on inclinations and the solid line is the \ibothros\ spectrum for a nearly edge-on inclination.}
\label{fig:spec_harm2_wscatt}
\end{figure}

We now apply \grm\ to the same {\tt HARM} simulation data used in the
\grm-\ibothros\ comparison above, this time with scattering enabled.
Figure~\ref{fig:spec_harm2_wscatt} shows the resulting spectrum with the
\ibothros\ result shown for comparison.  Evidently, scattering has
little effect on the sub-mm spectrum since the scattering optical depth is small ($\sim 10^{-4}$) but the model now predicts a
significant X-ray flux.  A more detailed analysis of Comptonized spectra
from GRMHD simulations in the context of Sgr A* will be given in a
separate paper \citep{M09}.

\begin{figure}
\plotone{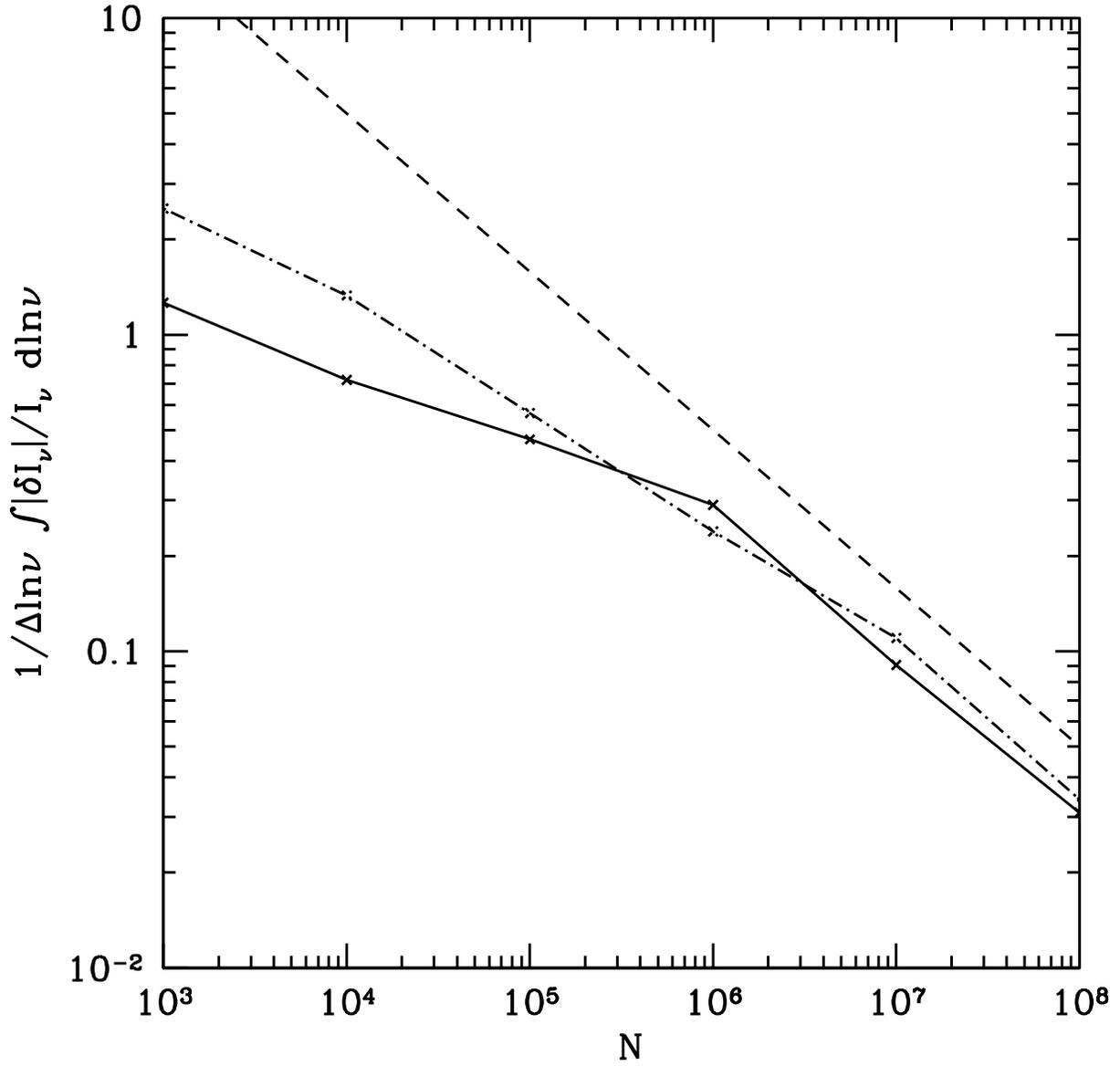}
\caption{Self-convergence test results for the spectra shown in Figure~\ref{fig:spec_harm2_wscatt}.  Convergence for the edge-on (solid) and face-on (dot-dash) spectra are shown.  The dashed line is proportional to $N^{-1/2}$.}
\label{fig:l1_full}
\end{figure}

Figure~\ref{fig:l1_full} shows the results of self-convergence tests for the full code.  Convergence is initially slow but quickly approaches a rate proportional to $N^{-1/2}$ as spectral bins become sufficiently sampled.  The spectra shown in Figure~\ref{fig:spec_harm2_wscatt}, which were taken from a single run with $10^9$ superphotons, were used for references.

\section{Summary}\label{SUMM}

We have described and tested a code that solves the radiative transfer
problem for optically thin ionized plasmas in general spacetimes.  The
code treats the full angular dependence of emission and absorption,
treats single Compton scattering exactly (double Compton and induced
Compton scattering are neglected), and can be easily adapted to simulate
emission from both analytic and numerical models.  While we have
specialized to synchrotron emission in this work, \grm\ is constructed
so that it is straightforward to include other relevant emission
mechanisms such as bremsstrahlung with only minimal modification.

As a demonstration of a practical use for \grm, we have computed the
first spectra, including synchrotron emission and Compton scattering,
from GRMHD models of a turbulent accretion disk.  Other potential
applications of our code are to neutron star accretion, emission from
relativistic blast waves, and any problem where relativistic bulk motion
makes radiation transfer treatments that expand the flow in orders of
$v/c$ problematic.

\acknowledgments

This work was supported by the National Science Foundation under grants
AST 00-93091, PHY 02-05155, and AST 07-09246, and by a Richard and
Margaret Romano Professorial scholarship, a Sony faculty fellowship, and
a University Scholar appointment to CFG.  Portions of this work were
performed while CFG was a Member at the Institute for Advanced Study in
academic year 2006-2007.  The authors are especially grateful to Shane
Davis and Stu Shapiro for their insights, and to Peter Goldreich, Fred
Lamb, and John Hawley for discussions.  We would also like to thank the anonymous referee for suggestions which helped improve this manuscript in clarity and completeness.

\newpage
\bibliographystyle{apj}
\bibliography{local}

\newpage

\newpage

\newpage

\newpage

\newpage

\newpage

\newpage

\newpage

\newpage

\newpage

\newpage

\newpage

\newpage

\newpage

\newpage

\newpage

\newpage

\end{document}